\documentclass{jfm}

\usepackage{graphicx}
\usepackage{subfigure}     
\usepackage{amssymb}
\usepackage{amsmath}
\usepackage{newtxtext}
\usepackage{newtxmath}
\usepackage{multirow}
\usepackage[sort&compress]{natbib}
\graphicspath{ {./images/} }
\usepackage{ifpdf}\ifpdf 
\usepackage[colorlinks,urlcolor=blue,citecolor=blue,linkcolor=blue,pdftex,hyperfigures,hypertexnames=false]{hyperref}
\else
\providecommand{\texorpdfstring}[2]{#1}
\fi
\makeatletter
\renewcommand{\thesubfigure}{\alph{subfigure}}
\renewcommand{\@thesubfigure}{(\thesubfigure)\hskip\subfiglabelskip}
\makeatother
\title{Liquid film dynamics with immobile contact line during meniscus oscillation}
\author{Xiaolong Zhang \and Vadim S. Nikolayev\corresp{\email{vadim.nikolayev@cea.fr}}}
\affiliation{Universit\'e Paris-Saclay, CEA, CNRS, SPEC, 91191 Gif-sur-Yvette Cedex, France}
\begin{document}
\maketitle
\begin{abstract}
This paper presents a theoretical analysis of the liquid film dynamics during the oscillation of a meniscus between a liquid and its vapour in a cylindrical capillary. By using the theory of Taylor bubbles, the dynamic profile of the deposited liquid film is calculated within the lubrication approximation accounting for the finiteness of the film length, i.e. for the presence of the contact line. The latter is assumed to be pinned on a surface defect and thus immobile; the contact angle is allowed to vary. The fluid flow effect on the curvature in the central meniscus part is neglected. This curvature varies in time because of the film variation and is determined as a part of the solution. The film dynamics depends on the initial contact angle, which is the maximal contact angle attained during oscillation. The average film thickness is studied as a function of system parameters. The numerical results are compared to existing experimental data and to the results of the quasi-steady approximation. Finally, the problem of an oscillating meniscus is considered accounting for the superheating of the capillary wall with respect to the saturation temperature, which causes evaporation. When the superheating exceeds a quite low threshold, oscillations with a pinned contact line are impossible anymore and the contact line receding caused by evaporation needs to be accounted for.
\end{abstract}

\section{Introduction}\label{Introduction}

The oscillating motion of menisci in thin capillaries is of importance for many applications. One can cite the liquid plugs that obstruct the airways in living organisms for certain pathologies \citep{Baudoin13}, the distribution of fluids in microfluidics \citep{Angeli08} and the oscillations caused by the vapour--liquid mass exchange in heat pipes. This latter application is targeted in the present work as it is relevant to different types of heat pipes. One can cite capillary pumped loops \citep{Zhang98} or loop heat pipes, where the pressure oscillations are observed \citep{Launay07} and impact the menisci in the capillary structure. The oscillation of menisci is of special importance for the pulsating heat pipes, called also oscillating heat pipes \citep{LauraATE17,EncycExp18,ATE21}, where the liquid films deposited by the oscillating liquid menisci as they recede provide the main channel of the heat and mass transfer. As the film evaporation rate is defined by the local film thickness, one needs to understand the film profile for adequate modelling of the heat pipe. The film evaporation description is the most challenging part because the liquid film can be partially dried out so that triple vapour-liquid-solid contact lines form. Strong heat and mass transfers occur in their vicinity \citep{EuLet12,Savva17} so the contact lines are important to model adequately.

The hydrodynamics of menisci has been extensively studied since the seminal articles of \citet{LL42}, \citet{Taylor} and \citet{Bretherton}. Since their works, a strong effort has been made to understand the dynamics of the Taylor bubbles (i.e. bubbles of the length larger than their diameter) and the liquid plugs that separate them. Originally, the hydrodynamics of such a process has been described theoretically within the creeping flow approximation (i.e. for vanishingly small Reynolds numbers) by using the lubrication approach for the liquid film description. The inertial effects have been accounted for by direct numerical simulation \citep{Talimi12}. In previous approaches, the liquid film was considered to be continuous, with no dry patches.

The objective of this work is twofold. First, we study the film created by the meniscus oscillation. Second, we want to understand the impact of the film edge, i.e. of the triple contact line, for the simplest case where it is pinned at a surface defect and is thus immobile.

The paper is structured as follows. After an introduction of the model in sec.~\ref{sec:Description}, the background theory of steady meniscus receding is briefly discussed in sec.~\ref{sec:LLfilm}. The meniscus oscillation is considered in sec.~\ref{sec:osctheor}. The theory is compared to two experimental works involving meniscus oscillation. While the main objective of this paper is to consider the oscillation with no heat and mass transfer, an interesting implication of these results for the film evaporation is discussed in sec.~\ref{oscevapsec}.

\section{Model description}\label{sec:Description}

\begin{figure}
  \centering
\includegraphics[width=0.6\textwidth]{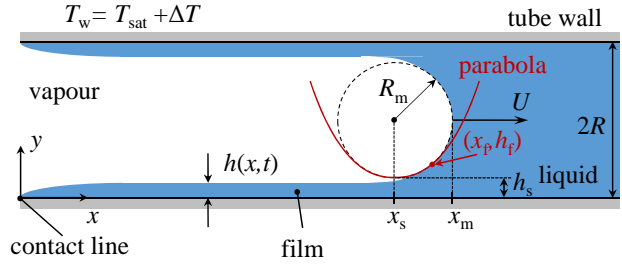}
  \caption{Sketch of the axial cross-section of a cylindrical capillary tube with the liquid film deposited by the receding meniscus. At $x=x_\mathrm{f}$, the film profile solution matches the right branch of the parabola (red solid line), which is a small-slope approximation of the circular meniscus shape (dashed line).}\label{fig:cross-section}
\end{figure}

Consider a cylindrical capillary tube of an inner radius $R$, containing a liquid and its vapour. The vapour--liquid interface is assumed to be axially symmetric (Fig.~\ref{fig:cross-section}).

The tube is assumed to be thin enough so the gravity force can be neglected. Following the classical approach \citep{Bretherton}, the vapour--liquid interface can be divided into the film and the meniscus regions. The liquid--vapour interface slope in the film region is assumed to be small so the film can be described with the lubrication theory. The meniscus region is assumed to be controlled by the surface tension only, thus being of constant curvature (shown in Fig.~\ref{fig:cross-section} with a circle of radius $R_\mathrm{m}$).

Because the vapour has a smaller density and viscosity compared with the liquid, the vapour pressure $p$ is assumed to be spatially homogeneous and the vapour-side viscous stress on the interface can be neglected. Under such assumptions, the lubrication theory results in the equation \citep{PF10} that describes the interface dynamics in the film region,
\begin{equation}
\label{eq:dimensional-GE}
\frac{\partial h}{\partial t}+\frac{\partial}{\partial x}\left( \frac{h^3}{3\mu}\frac{\partial\Delta p}{\partial x} \right) = -\frac{J}{\rho},
\end{equation}
where $h=h(x,t)$ is the local film thickness, and $J$ is the mass flux across the interface, defined to be positive at evaporation. Here, $\mu$ and $\rho$ are the liquid shear viscosity and density, respectively. The lubrication theory is applicable within the assumption $|\partial h/\partial x|\ll 1$. The pressure jump $\Delta p=p-p_\mathrm{l}$ (with $p_\mathrm{l}$, the liquid pressure) across the interface obeys the Laplace equation
\begin{equation}\label{eq:laplace}
  \Delta p=\sigma \left( K +\frac{1}{R - h} \right)\simeq \sigma\left( \frac{{\partial ^2}h}{\partial {x^2}}+\frac{1}{R_\mathrm{m}}\right),
\end{equation}
where $\sigma$ is the surface tension, $K$ is the two-dimensional interface curvature in the axial cross-section shown in Fig.~\ref{fig:cross-section}; in the small-slope approximation, $K\simeq\partial ^2h/\partial x^2$. Because of this limitation, such a film region theory is not able to describe the meniscus region. The radial contribution to the curvature $R_\mathrm{m}^{-1}$ is assumed to be independent of $x$ in the film region (where $h$ is much smaller than $R$). At a large $x$, Eq.~\eqref{eq:dimensional-GE} results in an increasingly larger $h$, where the viscous forces vanish and the surface tension alone controls the interface, so its curvature $\partial ^2h/\partial x^2$ becomes constant. Such a condition corresponds to a parabolic shape in the axial plane. This parabola needs to be joined to the circular meniscus, which results in the condition
\begin{equation}\label{eq:bc_LLrd}
\partial^2 h/\partial x^2|_{x=x_\mathrm{f}}=R_\mathrm{m}^{-1}.
\end{equation}
defined at the ending point $(x_\mathrm{f},\, h_\mathrm{f})$ of the film region.

\section{Isothermal problem with infinite film: steady solutions} \label{sec:LLfilm}

First, we would like to recall the theory for the case with no phase change ($J=0$), so Eq.~\eqref{eq:dimensional-GE} becomes
\begin{equation}\label{eq:GE}
\frac{\partial h}{\partial t} + \frac{\sigma}{3\mu} \frac{\partial}{\partial x}\left( h^3\frac{\partial ^3h}{\partial x^3} \right) =0.
\end{equation}
The velocity of the axial meniscus centre (assumed positive for a receding meniscus according to the $x$-axis direction choice) is denoted $U$. The contact line is not considered so the film is infinite. In this case, it is advantageous to choose the frame of reference linked to the moving meniscus where the axial coordinate becomes $x'=x-Ut$. Eq.~\eqref{eq:GE} can then be rewritten as
\begin{equation}
\label{eq:GEm}
\frac{\partial h}{\partial t} + \frac{\partial}{\partial x'}\left( \frac{\sigma }{3\mu}h^3 \frac{\partial ^3h}{\partial {x'}^3} -Uh \right) =0.
\end{equation}
This equation (and this frame of reference) is convenient to use in the present section and in Appendix~\ref{sec:relax} because the film is infinite there. However, the prime will be dropped to make notation less heavy. In all other sections, where the contact line is considered, the wall frame of reference will be used as it is more convenient, so the meaning of $x$ will be as in \eqref{eq:dimensional-GE} and \eqref{eq:GE}.

The steady version of Eq.~\eqref{eq:GEm} for the case of a constant positive velocity $U_\mathrm{r}$ (that should be used for $U$ in this case) is the Landau--Levich equation \citep{LL42} describing the flat infinite film being deposited by a receding meniscus. Note that the film in a cylindrical capillary \citep{Bretherton} is described by the same equation because of the approximation \eqref{eq:laplace}.

The boundary conditions at $x\to-\infty$ describe the flat film of the yet unknown thickness $h_\mathrm{r}$,
\begin{equation}\label{eq:bc_LLl}
h=h_\mathrm{r},\; \partial h/\partial x=0.
\end{equation}
\begin{table}
\renewcommand\arraystretch{1.5}
\centering
\begin{tabular}{lccc}
\multirow{2}{*}{variable} & \multirow{2}{*}{notation} & \multicolumn{2}{c}{reference values used in} \\
\cline{3-4}
&  &  Sec. \ref{sec:LLfilm} and Appendix \ref{sec:relax} & Other sections \\
axial coordinate      & $x$      & $h_\mathrm{r}Ca_\mathrm{r}^{-1/3}$   & $\alpha R Ca_0^{1/3}$ \\
film thickness        & $h$      & $h_\mathrm{r}$     & $\alpha R Ca_0^{2/3}$ \\
time                  & $t$      & $h_\mathrm{r} Ca_\mathrm{r}^{-1/3} U_\mathrm{r}^{-1}$  & $\alpha R Ca_0^{1/3}U_0^{-1}$ \\
velocity              &$U$       & $U_\mathrm{r}$        & $U_0=2\pi A/P$    \\
\end{tabular}
\caption{Dimensional reference values that are used to make the governing equations non-dimensional.}\label{tab:references}
\end{table}

The scaling of this problem (Table~\ref{tab:references}) is based on $h_\mathrm{r}$. The characteristic axial length scale involves the capillary number $Ca_\mathrm{r}=\mu U_\mathrm{r}/\sigma$ and is chosen in such a way that the dimensionless equation
\begin{equation}
\label{eq:GEm-dimensionless}
\frac{\partial }{\partial \tilde x}\left( { \frac{{\tilde h}^3}{3}\frac{{{\partial ^3}\tilde h}}{{\partial {\tilde x ^3}}}} - \tilde h \right) =0,
\end{equation}
does not contain any constants; the tilde means hereafter the corresponding dimensionless variable.

By integrating Eq.~\eqref{eq:GEm-dimensionless} from $-\infty$ to $\tilde x$ and using the conditions~\eqref{eq:bc_LLl} (in the dimensionless form, $\tilde h(\tilde x\to-\infty)=1$), one obtains
\begin{equation}\label{eq:LL}
\frac{\partial^3\tilde h}{\partial\tilde x^3} =3\frac{\tilde h-1}{\tilde h^3},
\end{equation}
which is equivalent to the \citeauthor{Bretherton} equation within a factor 3 that we leave in the equation instead of putting inside the scaling parameters of Table ~\ref{tab:references}. This helps us to avoid it in many formulas used below.

Consider now the behaviour at large $\tilde x$. Eq.~\eqref{eq:LL} remains valid until the transition region (between the film and the meniscus), in which $1\ll\tilde h <R$. From Eq.~\eqref{eq:LL}, ${\partial^3\tilde h}/{\partial\tilde x^3} \simeq 0$. This means that, at large $\tilde x$, the second derivative is finite, which is compatible to the condition \eqref{eq:bc_LLrd}. Eq.~\eqref{eq:LL} can be integrated numerically (see \citet{Nikolayev14} for details). The ending point $(\tilde x_\mathrm{f},\,\tilde h_\mathrm{f})$ of the integration interval is chosen from a condition that $\partial^2\tilde h/\partial\tilde x^2$ reaches a plateau (with a required accuracy). Such a calculation results in a numerical value for this plateau
\begin{equation}\label{eq:bc_LLr}
\left.\frac{\partial^2\tilde h}{\partial\tilde x^2}\right|_{\tilde x=\tilde x_\mathrm{f}}=\alpha\simeq 1.3375
\end{equation}
The resulting profile $\tilde h(\tilde x)$ can be found in Fig.~\ref{fig:film-rlx} of Appendix \ref{sec:relax}. Note that $\alpha$ is equivalent to the numerical value 0.643 originally found by \citet{Bretherton}; $\alpha\simeq 0.643\cdot 3^{2/3}$, where the factor appears because of the different scaling. By comparing Eqs.~(\ref{eq:bc_LLrd}, \ref{eq:bc_LLr}), one finds the expression for the film thickness
\begin{equation}\label{eq:hinf}
 h_\mathrm{r}=\alpha R_\mathrm{m}Ca_\mathrm{r}^{2/3}.
\end{equation}

The only yet unknown quantity is $R_\mathrm{m}$ that can be found as proposed by \citet{Klaseboer14}. As mentioned before, near the point $(\tilde x_\mathrm{f},\,\tilde h_\mathrm{f})$ the film shape should satisfy the condition~\eqref{eq:bc_LLrd} which means that at $h\gg h_\mathrm{r}$ it asymptotically approaches a parabola
\begin{equation}\label{eq:par}
y=(x-x_\mathrm{s})^2/(2R_\mathrm{m})+h_\mathrm{s},
\end{equation}
where the parameters $(x_\mathrm{s},\,h_\mathrm{s})$ are yet to be determined. To understand their meaning, one recalls that near its minimum where its curvature is $R_\mathrm{m}^{-1}$, the parabola approximates the circular meniscus profile
\begin{equation}\label{circ}
(x-x_\mathrm{s})^2+(y-h_\mathrm{s}-R_\mathrm{m})^2=R_\mathrm{m}^2.
\end{equation}
It is evident now that $(x_\mathrm{s},\,h_\mathrm{s})$ is the circle lowest point. One can obtain $(x_\mathrm{s},\,h_\mathrm{s})$  by fitting the film profile $\tilde h(\tilde x)$ near the point $(\tilde x_\mathrm{f},\,\tilde h_\mathrm{f})$ to a parabola. \citet{Klaseboer14} report that the dimensionless value of $\tilde h_\mathrm{s}$ slightly grows with $\tilde h_\mathrm{f}$.  \citet{Bretherton} gives $\tilde h_\mathrm{s}=2.79$. The asymptotic value $\tilde h_\mathrm{s}=2.90$ is obtained for $\tilde h_\mathrm{f}> 10^6$. However, to obtain a continuous overall interface profile, the matching point $(\tilde x_\mathrm{f},\,\tilde h_\mathrm{f})$ film--parabola should be lower than the point where the parabola--circle transition occurs. This requires $h_\mathrm{f}< R$. For the  value $\tilde h_\mathrm{f}\simeq 50$ that satisfies this condition in practical situations, \citet{Klaseboer14} find $\tilde h_\mathrm{s}=2.5$, which is also the value found from the experimental data fits, as discussed below. In summary, the $\tilde h_\mathrm{s}$ variation is weak and one can consider that the circle is nearly invariant of the specific $\tilde h_\mathrm{f}$ choice.

\citet{Klaseboer14} have proposed the equation
\begin{equation}
R_\mathrm{m}+h_\mathrm{s}=R\label{eq:Rmeq}
\end{equation}
that centres the circle with respect to the tube and thus links $R_\mathrm{m}$ to $R$, cf. Fig.~\ref{fig:cross-section}. By using Eq.~\eqref{eq:hinf} in this equation, one finally obtains
\begin{equation}\label{eq:crv-receding}
R_\mathrm{m}= \frac{R}{1 + \alpha \tilde h_\mathrm{s} Ca_\mathrm{r}^{2/3}}.
\end{equation}
By combining Eqs.~(\ref{eq:hinf}, \ref{eq:crv-receding}), one can now finalise the film thickness expression
\begin{equation}\label{eq:hr}
  h_\mathrm{r}=\frac{\alpha R Ca_\mathrm{r}^{2/3}}{1 + \alpha \tilde h_\mathrm{s} Ca_\mathrm{r}^{2/3}}.
\end{equation}

The value $\tilde h_\mathrm{s}\simeq 2.5$ has been determined by \citet{Aussillous} from the experimental data fits. For $Ca_\mathrm{r}\to 0$, $R_\mathrm{m}\simeq R$ and Eq.~\eqref{eq:hr} reduces to \citeauthor{Bretherton}'s original expression
\begin{equation}\label{Breth}
h_\mathrm{r}=\alpha R Ca_\mathrm{r}^{2/3}.
\end{equation}

One can consider the meniscus advancing at a constant velocity $U=-U_\mathrm{a}$ (where $U_\mathrm{a}$ is the modulus of the advancing velocity) over the pre-existing film of thickness $h_\mathrm{r}$. Such a motion has been understood as well. It has been shown \citep{Bretherton} that the film has a wavy shape (ripples) near the meniscus, cf. the solid curve in Fig.~\ref{fig:film-rlx} of Appendix \ref{sec:relax}. The wavelength of ripples depends on the ratio $U_\mathrm{a}/U_\mathrm{r}$ \citep{Maleki11,Nikolayev14}, where $U_\mathrm{r}$ can be deduced (with Eq.~\ref{Breth}) from $h_\mathrm{r}$. The meniscus radius for the steady advancing case was determined with Eq.~\eqref{eq:Rmeq} by \citet{Cherukumudi15}.

\section{Oscillations in the presence of a contact line}\label{sec:osctheor}

A previous study \citep{Nikolayev14} demonstrates the film behaviour for the case of an infinite film. There is no physical criterion imposing its thickness so it is another independent parameter. When the meniscus approaches the leftmost (in the reference of Fig.~\ref{fig:cross-section}) position observed during oscillations, the ripples created near the advancing meniscus propagate over the film to infinity, so there is no possible periodical regime. This propagation is amplified by the discrepancy between the imposed film thickness and the film thickness \eqref{eq:hinf} defined by the receding meniscus velocity, which is zero at the leftmost point. Thus a discrepancy exists for any imposed film thickness. In practical situations of oscillating motion \citep{PRF16,Rao17}, the contact line appears because of the film evaporation caused by the tube wall heating. The film completely vaporises beyond the leftmost meniscus position. Before addressing the heating case, in this section we discuss the film shape in the presence of contact line without any heating.

The hydrodynamics of the pinned (static) contact line is simpler than the dynamic case. For this reason one needs to understand it first. This is a purpose of this work. The contact line pinning often occurs in capillaries \citep{Mohammadi15}. It is caused by the wall heterogeneity (surface defects) that can be either chemical or geometrical (surface roughness). The heterogeneity can be modelled as a spatial variation of surface energy. The result of such a theory \citep{PRE14} is that the microscopic contact angle averaged along the contact line can vary between the static advancing $\theta_\textrm{adv}$ and the static receding $\theta_\textrm{rec}$ angles while the contact line remains immobile. In our calculation, $\theta_\textrm{adv} - \theta_\textrm{rec}$ (called wetting hysteresis) is assumed to be sufficiently large so the contact line always remains immobile. In experiments, the hysteresis can be as large as $50^\circ$ \citep{deG}, which is larger than the angle oscillation magnitude considered below.

At oscillations with the fixed contact line, there are no vortices near it \citep{Ting} and the flow is known to be well described by the lubrication approximation, even down to the nanometric scale \citep{Mortagne17}.

\subsection{Relaxing the pressure divergence by the Kelvin effect}\label{RelaxSec}

The Stokes problem of the straight wedge with a varying opening angle leads to the logarithmic pressure divergence, cf. Appendix \ref{MoffatSec}. Such a divergence is integrable and thus does not cause a paradox similar to that of the moving contact line. However, the infinite pressure is non-physical. In addition, the pressure boundary condition at the contact line would be difficult to use in calculation because it requires prior knowledge of the contact angle and its time derivative (cf. Eq.~\ref{plrl}) that need to be determined themselves during the solution procedure. As we consider volatile fluids, the phase change together with the Kelvin effect are introduced. The latter makes the pressure to be finite everywhere, as shown below. The problem is formulated here for a general case where the tube wall can be superheated or subcooled with respect to the saturation temperature $T_\mathrm{sat}$ corresponding to the imposed vapour pressure $p$. The wall superheating is denoted $\Delta T$. The tube wall temperature is thus $T_\mathrm{w}=T_\mathrm{sat}+\Delta T$.

Conventional hypotheses concerning the liquid film mass exchange \citep{PF10} are applied. A linear temperature profile in the radial direction is assumed in the thin liquid film, so the energy balance at the interface results in the mass flux
\begin{equation}\label{eq:J}
  J = \frac{k(T_\mathrm{w}-T_\mathrm{int})}{h\cal L},
\end{equation}
where $T_\mathrm{int}$ is the temperature of the vapour--liquid interface, $k$ is the liquid heat conductivity and $\cal L$ is the latent heat. The film is assumed here to be thin with respect to $R$ so the one-dimensional conduction description applies. The evaporation impact on a film is twofold. First, the film thickness decreases with time everywhere along the film, which is described by the balance of the first and the right-hand side term of Eq.~\eqref{eq:dimensional-GE}. Generally, the film thinning is not strong during an oscillation period \citep{LauraATE17}.

We focus here on the second effect that appears because of the strength of evaporation in a narrow vicinity of contact line. If the vapour--liquid interface was at a fixed saturation temperature ($T_\mathrm{int}=T_\mathrm{sat}$), the mass flux $J$ \eqref{eq:J} would diverge at the contact line $h=0$ as $J\sim \Delta T/h$, which is non-physical because total evaporated mass (the integral of $J$) would be infinite.

The Kelvin effect, i.e. the dependence of $T_\mathrm{int}$ on the interfacial pressure jump $\Delta p$
\begin{equation}\label{eq:Ti}
T_\mathrm{int} = T_\mathrm{sat}\left( {1 + \frac{\Delta p}{\cal L \rho}} \right)
\end{equation}
can relax the singularity \citep{EuLet12}, because it allows $T_\mathrm{int}$ to vary along the interface so it can attain the wall temperature $T_\mathrm{w}$ at the contact line so the mass flux
\begin{equation}\label{J0}
  J(x\to 0)=0.
\end{equation}
From the temperature continuity, one obtains the condition
\begin{equation}\label{eq:p_cl}
\Delta p(x \to 0) =\Delta p_\mathrm{cl},
\end{equation}
where a constant pressure jump at the contact line is introduced as
\begin{equation}\label{Dpcl}
\Delta p_\mathrm{cl}=\frac{{\cal L}{\rho}}{T_\mathrm{sat}} \Delta T.
\end{equation}
One can show that a solution that satisfies this condition can indeed be found (cf. Appendix \ref{LubAsSec}).

Eqs.~(\ref{eq:J}, \ref{eq:Ti}) result in
\begin{equation}\label{J}
 J = \frac{k}{h\cal L}\left(\Delta T- \Delta p\frac{ T_\mathrm{sat}}{\cal L \rho} \right).
\end{equation}
With its substitution into Eq.~\eqref{eq:dimensional-GE}, the governing equation becomes
\begin{equation}
\label{eq:GE-kelvin}
\frac{\partial h}{\partial t} + \frac{\partial }{\partial x}\left(\frac{h^3}{3\mu} \frac{\partial\Delta p}{\partial x}\right)=\frac{\Delta p-\Delta p_\mathrm{cl}}{h}\frac{kT_\mathrm{sat}}{ ({\cal L} \rho)^2}.
\end{equation}
The problem is now regular (because $\Delta p$ is not divergent anymore), unlike other microscopic approaches \citep{Savva17}. As the Kelvin effect alone is capable of relaxing the contact line singularity, the other microscopic scale effects such as hydrodynamic slip, Marangoni effect or interfacial kinetic resistance \citep{EuLet12} are not crucial anymore. They are not included in our model for the sake of clarity.

The characteristic size of the contact line vicinity where the Kelvin effect is important is $\ell_\mathrm{K}$ \eqref{lK}, cf. Appendix \ref{LubAsSec} for more details. It is nanometric \citep{PRE13movingCL} and is thus significantly smaller than the characteristic scale of film shape variation that we call macroscopic. For this reason, Eq.~\eqref{eq:GE-kelvin} can be understood within a multi-scale paradigm in the spirit of the asymptotic matching techniques \citep{PRE13movingCL}. In the inner region, commonly called the microregion, the first (transient) term is negligible with respect to the Kelvin term ($\Delta p$ containing term in the r.h.s.). The problem is reduced to that of Appendix \ref{sec:app}. To summarise it, when $\Delta T\neq 0$, a strong interfacial curvature that exists in the microregion can cause a difference between the microscopic contact angle $\theta_\mathrm{micro}$ and the interface slope $\theta$ defined at $x\to\infty$ within the microregion. In the outer (macroscopic) region, the Kelvin effect is negligible so the $\Delta p$ term on the r.h.s. of Eq.~\eqref{eq:GE-kelvin} vanishes. For $\Delta T=0$, this equation is that of the isothermal problem \eqref{eq:GE}. Both problems can be matched at scale $x_\mathrm{meso}\gg \ell_\mathrm{K}$, much smaller than the macroscopic scale. The apparent contact angle visible on this latter scale is thus equal to $\theta$. It is assumed hereafter that the pinning occurs at a length scale smaller than $\ell_\mathrm{K}$, i.e. there are nanometric defects with sharp borders on which the contact line is pinned.



In sec. \ref{sec:osctheor}, a globally isothermal problem is considered, $T_\mathrm{w}=T_\mathrm{sat}$, so $\Delta T=0$ and $\Delta p_\mathrm{cl}=0$, which means $\theta=\theta_\mathrm{micro}$. At such conditions, the mass exchange appearing at the macroscopic scale is very weak so that it can be safely neglected. This does not mean, however, that the mass exchange is absent in the microregion where $\Delta p$ can be large, as mentioned above. The mass flux $J$ scales with $\Delta p$ according to Eq.~\eqref{J}, so phase change occurs. The situation here shares certain similarities with the contact line motion paradox solved by the Kelvin effect \citep{PRE13movingCL}. Consider e.g. an increasing in time $\theta$. According to Eq.~\eqref{powAs}, $\Delta p>0$ in the very contact line vicinity so the condensation occurs there. It is compensated exactly by evaporation farther away from the contact line, so the net mass exchange is zero. It should be noted that the fluid flow associated with the phase change is strongly localised within a nanoscale distance from the contact line comparable to $\ell_\mathrm{K}$.

In conclusion, all results obtained in the present sec. \ref{sec:osctheor}, can be seen as obtained with Eq.~\eqref{eq:GE} because the microregion details cannot be resolved at the macroscale pictured in the figures below. However, the numerical calculations of the regularised Eq.~\eqref{eq:GE-kelvin} are carried out in reality.

\subsection{Oscillation problem statement}

The meniscus now oscillates, and the position $x_\mathrm{m}$ of its centre (which is the experimentally measurable quantity) travels periodically with a period $P$ and an amplitude $A$. One can assume its harmonic oscillation
\begin{equation}\label{xm}
x_\mathrm{m} (t)= x_\mathrm{i} +  A[ 1 - \cos(2\pi t/P)],
\end{equation}
where $x_\mathrm{i}$ is the initial meniscus centre position. Alternatively, one can take the experimentally measured dependence $x_\mathrm{m}(t)$ while comparing the data with the experiment (cf. sec.~\ref{expqual} below). The contact line is pinned at the position $x=0$, and the contact angle $\theta$ varies. For the harmonic oscillation case, the meniscus velocity is $U(t)= U_0\sin(2\pi t/P)$, where the velocity amplitude $U_0=2\pi A/P$ is convenient to choose as the characteristic velocity to define the capillary number $Ca_0=\mu U_0/\sigma$ and to make all the quantities dimensionless (cf. Table ~\ref{tab:references}).

Because of the fixed contact line, the frame of reference of the tube wall is chosen. Eq.~\eqref{eq:GE-kelvin} (with the substitution of Eq.~\ref{eq:laplace}) is solved for $ x\in[0, x_\mathrm{f}]$. The length $x_\mathrm{f}$ is imposed as explained in secs.~\ref{CurvSec},~\ref{NumSec} below.

The boundary conditions are defined as
\begin{subequations}\label{eq:BC-t}
\begin{align}
h\left(x = 0\right) &= 0, \label{bc0}\\
  \left.\frac{\partial \Delta p}{\partial  x} \right|_{ x \to 0} &= 0, \label{bc1} \\
\Delta p\left( x =  x_\mathrm{f}\right) &= R_\mathrm{m}^{-1},\label{bcf}\\
 h\left( x =  x_\mathrm{f}\right) &=  h_\mathrm{f},\label{bc2}
\end{align}
\end{subequations}
where $R_\mathrm{m}$ and $h_\mathrm{f}$ are discussed in sec.~\ref{CurvSec}. The condition \eqref{bc0} is a geometrical constraint at the contact line. Eq.~\eqref{bc1} is a weaker form of the condition \eqref{eq:p_cl} used to provide numerical stability. The boundary conditions (\ref{bcf}, \ref{bc2}) impose the liquid film curvature and thickness at the right end of the integration interval for each time moment.

\subsection{Determination of the meniscus curvature}\label{CurvSec}

For a small film thickness, one can assume that the meniscus radius $R_\mathrm{m}$ is constant and equal to $R$ during oscillation. It is actually a good approximation for a small $Ca_0\lesssim 10^{-3}$. However, at a larger $Ca_0$, the film thickness impacts $R_\mathrm{m}$ (see sec.~\ref{CurvResSec} below). Since the film thickness depends on the meniscus velocity, so does the meniscus radius $R_\mathrm{m}$, cf. sec.~ \ref{sec:LLfilm}. Therefore, $R_\mathrm{m}$ varies in time. In this section, we generalise to any meniscus dynamics the method for $R_\mathrm{m}$ determination \citep{Klaseboer14} discussed above for the steady receding case.

Similarly to the steady case of sec.~\ref{sec:LLfilm}, one needs first to match the film shape $h(x)$ to the parabola \eqref{eq:par}, and then the parabola to a circle \eqref{circ}. The matching between the film and the parabola means both the continuity and the smoothness (equality of the derivatives)
\begin{subequations}\label{match}
\begin{align}
&h_\mathrm{f}=(x_\mathrm{f}-x_\mathrm{s})^2/(2R_\mathrm{m})+h_\mathrm{s},\label{hs}\\
&\left.\frac{\partial h}{\partial x} \right|_{ x= x_\mathrm{f}}= \frac{x_\mathrm{f}- x_\mathrm{s}}{R_\mathrm{m}},
\end{align}
where the left-hand sides come from the film calculation and all the parabola parameters are time dependent. As in the approach of \citet{Klaseboer14}, Eq.~\eqref{eq:Rmeq} serves to find $R_\mathrm{m}$. We introduce in addition a relationship of the abscissas of the lowest and rightmost points of a circle that is needed to define $x_\mathrm{s}$ (Fig.~\ref{fig:cross-section}):
\begin{equation}
R_\mathrm{m}+ x_\mathrm{s}=x_\mathrm{m}.\label{xseq}
\end{equation}
\end{subequations}

In the present algorithm, $x_\mathrm{f}$ imposed to such a value that the difference $x_\mathrm{m}-x_\mathrm{f}$ does not vary in time and $h_\mathrm{f}=h(x_\mathrm{f})$ remains large with respect to the deposited film thickness. As discussed in sec.~\ref{sec:LLfilm}, the solution is nearly independent of the specific choice of $h_\mathrm{f}$  (and thus of $x_\mathrm{f}$). The set of Eqs.~\eqref{eq:laplace}, \eqref{eq:Rmeq} and \eqref{eq:GE-kelvin}--\eqref{match} is then complete, so the film shape and the unknown parameters ($R_\mathrm{m},\,x_\mathrm{s},\, h_\mathrm{s},\, h_\mathrm{f}$) can be determined for each $t$.

\subsection{Initial conditions and solution periodicity}\label{PerSec}

One needs to define now the initial film shape $h(x,0)$ at $t =0$, which corresponds to the (yet unspecified) leftmost meniscus position $x_\mathrm{i}$ according to Eq.~\eqref{xm}. As an initial film profile, we choose that of equilibrium satisfying the condition $\partial h/\partial t=0$ that can be used in  Eq.~\eqref{eq:GE}. From the boundary condition \eqref{bc1}, one finds straightforwardly $\partial^3 h/\partial x^3=0$, i.e. the parabolic shape
\begin{subequations}\label{curvanglei}
\begin{equation}\label{init}
  h(x,0)=\frac{x^2}{2R_\mathrm{m,i}}+\theta_\mathrm{i} x,
\end{equation}
where $\theta_\mathrm{i}\equiv\theta(t=0)$ is the initial contact angle. It serves as another boundary condition, additional to \eqref{bc0} and \eqref{bcf}. By applying Eqs.~\eqref{eq:Rmeq} and \eqref{match} at $t=0$, one gets
\begin{align}
R_\mathrm{m,i}\equiv R_\mathrm{m}(t=0)&=R/(1-\theta_\mathrm{i}^2/2), \label{eq:curvi}\\
x_\mathrm{i}&=R_\mathrm{m,i}(1-\theta_\mathrm{i}).\label{eq:xi}
\end{align}
\end{subequations}
These expressions are the small-angle approximations of the expressions $R_\mathrm{m,i}=R/\cos\,\theta_\mathrm{i}$ and $x_\mathrm{i}=R_\mathrm{m,i}(1-\sin\theta_\mathrm{i})$ because Eq.~\eqref{init} is an approximation of the initially spherical meniscus.

During oscillation, the liquid is driven by the meniscus motion and the free interface remains in the state where $\partial h/ \partial t$ is always balanced by the curvature gradient, more precisely, by the second term of Eq.~\eqref{eq:GE}. This occurs because there are no other forces, in particular, no inertia. When the meniscus comes near the leftmost position, $U$ decreases and the system approaches the state \eqref{curvanglei} with no curvature gradient, thus, $| \partial h/ \partial t| $ decreases to zero or almost zero. It is not a rigorous proof that the state~\eqref{curvanglei} belongs to the limit cycle of the system, albeit, it should be quite close to it. This is surely true when the relaxation time $t_\textrm{rel} \ll P$, which is our case (cf. Appendix~\ref{sec:relax} for $t_\textrm{rel}$ discussion). Indeed, the numerical simulations show that $h(x,P)$ is indistinguishable from $h(x,0)$, cf. Fig.~\ref{fig:P=50_hx_curves} below. So do all other parameters (curvature, contact angle, etc.). This finding allows us to simulate a unique period.

\subsection{Numerical implementation}\label{NumSec}

The scales for the main quantities to make them dimensionless are shown in Table~\ref{tab:references}. With such a ``natural'' scaling three main dimensionless parameters are left: $\theta_\mathrm{i}$, $\tilde P$ and $Ca_0$. All quantities will be studied in this parametric space. The dimensionless amplitude is linked to the period
\begin{equation}\label{eq:AP}
  \tilde A= \tilde P/(2\pi),
\end{equation}
where a dimensionless quantity is denoted with a tilde. There is one more dimensionless parameter
\begin{equation}\label{N}
 N=\frac{\mu k T_\mathrm{sat}}{{\cal L}\rho\alpha R \,Ca_0}
\end{equation}
that describes the magnitude of the Kelvin effect in the microregion. However, it does not impact the interface shape at the film scale provided the characteristic microscopic scale \eqref{lK} is chosen to be small enough (cf. sec.~\ref{RelaxSec}). The mesh size is exponentially refined near the contact line (as $\tilde{x}\to 0$) to capture the contact angle variation without considerably increasing the total number of nodes \citep{PF10}.

At $\tilde t=0$, a value of $\tilde x_\mathrm{f}=10$, for which $\tilde h_\mathrm{f}$ is around 50, cf. the discussion in sec.~\ref{sec:LLfilm}. At $\tilde t>0$, the difference $\tilde x_\mathrm{m}-\tilde x_\mathrm{f}$ is maintained constant, and equal to that at $\tilde t=0$. To avoid the discretisation error for the contact angle, the initial interface profile is determined numerically by solving the equilibrium version of Eq.~\eqref{eq:GE-kelvin}, instead of using the analytical profile \eqref{init}.

Eq.~\eqref{eq:GE-kelvin} is solved numerically with the finite volume method (FVM), which is more stable numerically \citep{Patankar} than a more conventional finite difference method. In one dimension, a finite volume is just a segment. The variables such as $h$ and their even-order derivatives are defined at its centre (called a node), while the odd-order derivatives are defined at the segment ends. The FVM has the advantage that the liquid flux is continuous at the segment ends. Nonlinear terms are managed by iteration: they include values from the previous iteration. The numerical algorithm is similar to that used by \citet{PF10}.

One is interested in amplitudes that are large with respect to the meniscus width $x_\mathrm{m}-x_\mathrm{s}$, which means large dimensionless periods of oscillation, see Eq.~\eqref{eq:AP}. This signifies that the computational domain size varies considerably during oscillations. The grid thus needs to be adaptive and the calculation time can be of the order of a day on a regular PC.

\subsection{Interface profile during oscillation}\label{liqprof}

\begin{figure}
  \centering
  \includegraphics[width=0.5\textwidth]{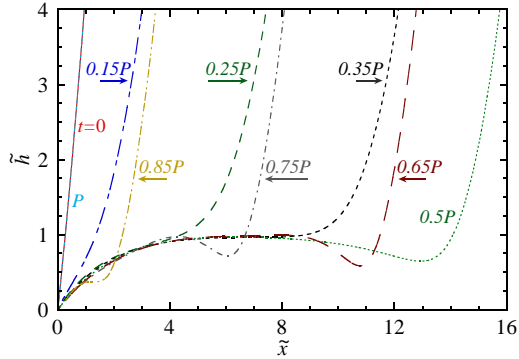}
  \caption{Periodic interface shape variation at oscillation for $\tilde P=50$, $Ca_0=10^{-3}$ and $\theta_\mathrm{i}=20^\circ$. The labels give the times corresponding to each profile, and arrows indicate the meniscus motion direction.}\label{fig:P=50_hx_curves}
\end{figure}

Figure~\ref{fig:P=50_hx_curves} shows the interface profiles at several time moments during oscillation. The meniscus motion follows the harmonic law \eqref{xm}. The liquid film is deposited until $t=P/2$. For $t>P/2$, the meniscus advances over the deposited film. The ripples near the meniscus appear, like during the steady meniscus advance discussed in sec.~\ref{sec:LLfilm}. The interface profiles $\tilde h(\tilde x,0)$ and $\tilde h(\tilde x,\tilde P)$ are indistinguishable, which confirms the periodicity of the oscillations.

\begin{figure}
  \centering
  \subfigure[ ]{\includegraphics[width=0.48\textwidth]{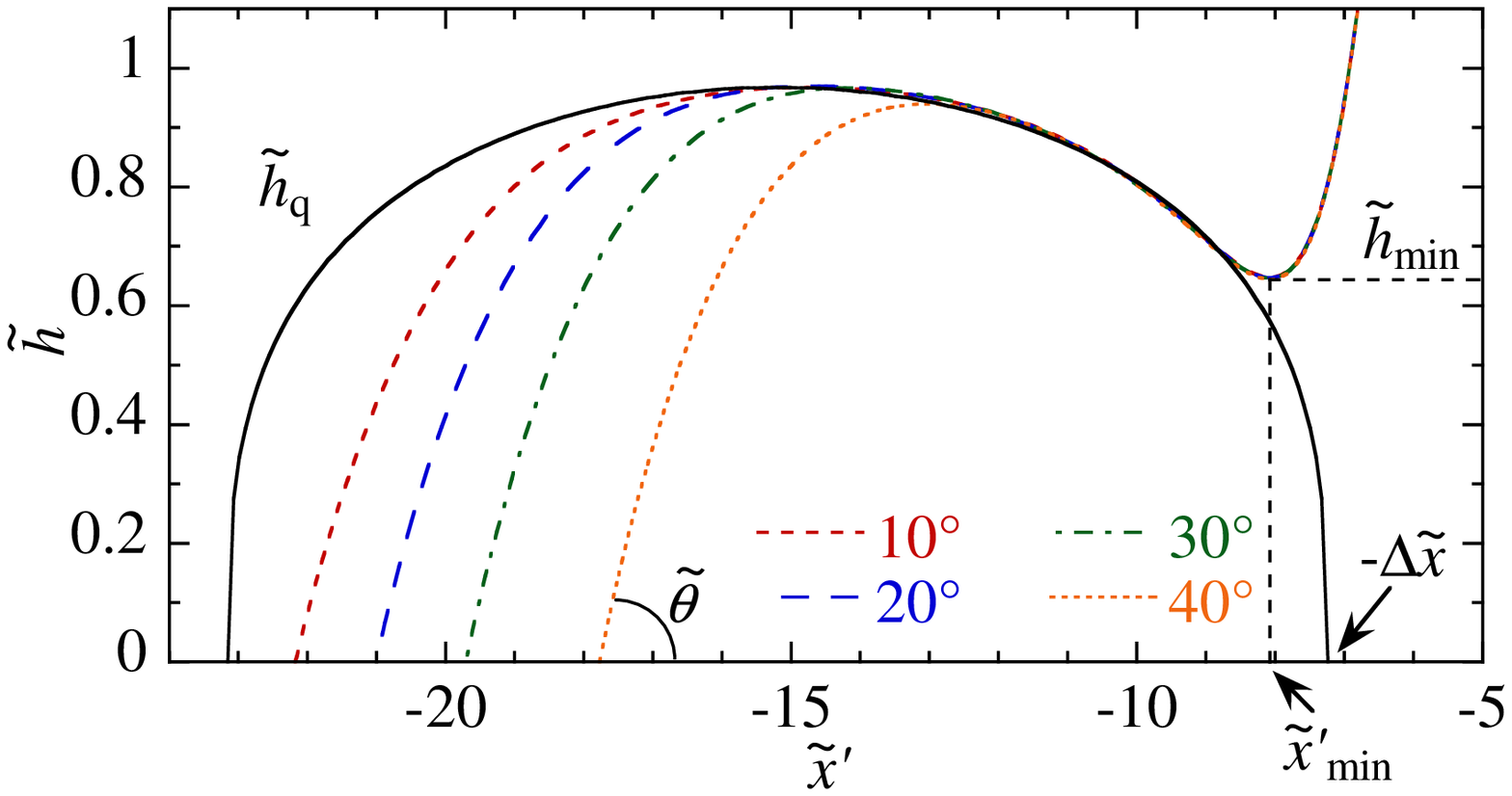}\label{P=50_h_vs_x}}\hfill
  \subfigure[ ]{\includegraphics[width=0.48\textwidth]{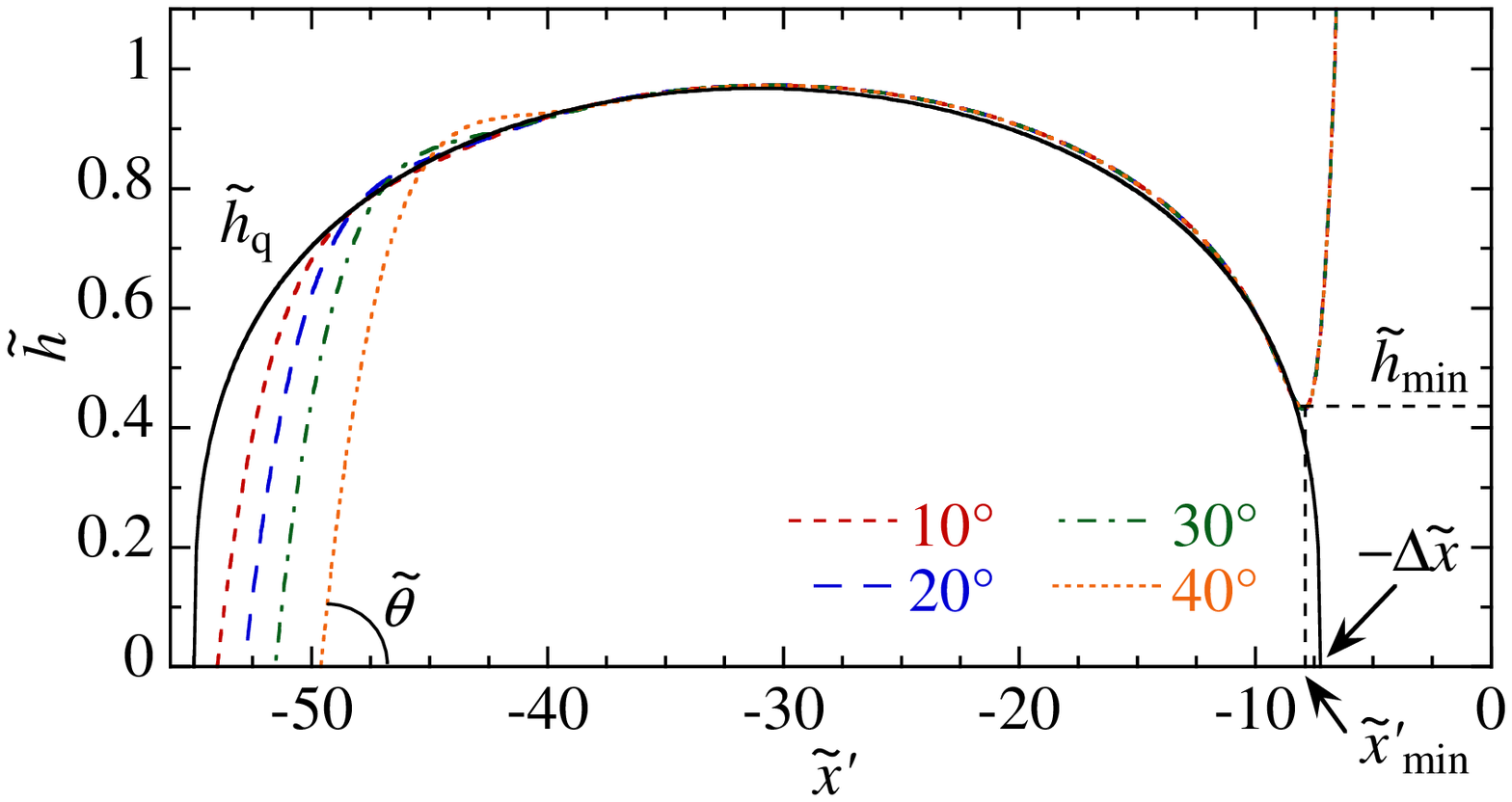}\label{P=150_h_vs_x}}
  \caption{Film shapes in the meniscus centre reference at $t=P/2$ for $Ca_0=10^{-3}$ and different  $\theta_\mathrm{i}$ and $P$. The quasi-steady profiles $\tilde h_\mathrm{q}$ discussed in sec.~\ref{QuasiSec} are shown for comparison. The scaled contact angle $\tilde{\theta}$ defined as $\tan\tilde{\theta}=Ca_0^{-1/3}\tan\theta$ is indicated; (a) $\tilde P=50$ and (b) $\tilde P=150$.}\label{fig:film-CAs}
\end{figure}

Fig.~\ref{fig:film-CAs} shows examples of interface profiles at $t=P/2$ (when the meniscus is at its rightmost position so the film length attains its maximum) for several values of the initial contact angle $\theta_\mathrm{i}$. All the film profiles are presented in the meniscus reference
\begin{equation}\label{xprime}
x'=x-2A-x_\mathrm{i},
\end{equation}
meaning that the meniscus centre is at $x'=0$, cf. Eq.~\eqref{xm}. One can see that the interface shape near the meniscus is independent of $\theta_\mathrm{i}$. During the meniscus receding, the film loses information about the contact line, so the film shape near the meniscus is controlled by the meniscus dynamics only. This is not surprising as the flow in the film is expected to occur only in the contact line and meniscus vicinities, but not in the middle of the film. The film profiles exhibit a local minimum $h_\mathrm{min}$ discussed in sec.~\ref{ContVarSec} below. It appears because of the meniscus velocity reduction at the end of a half-period.

\subsection{Contact angle during oscillation}\label{ContVarSec}

\begin{figure}
 \centering
 \subfigure[ ]
  {\includegraphics[width=0.45\textwidth]{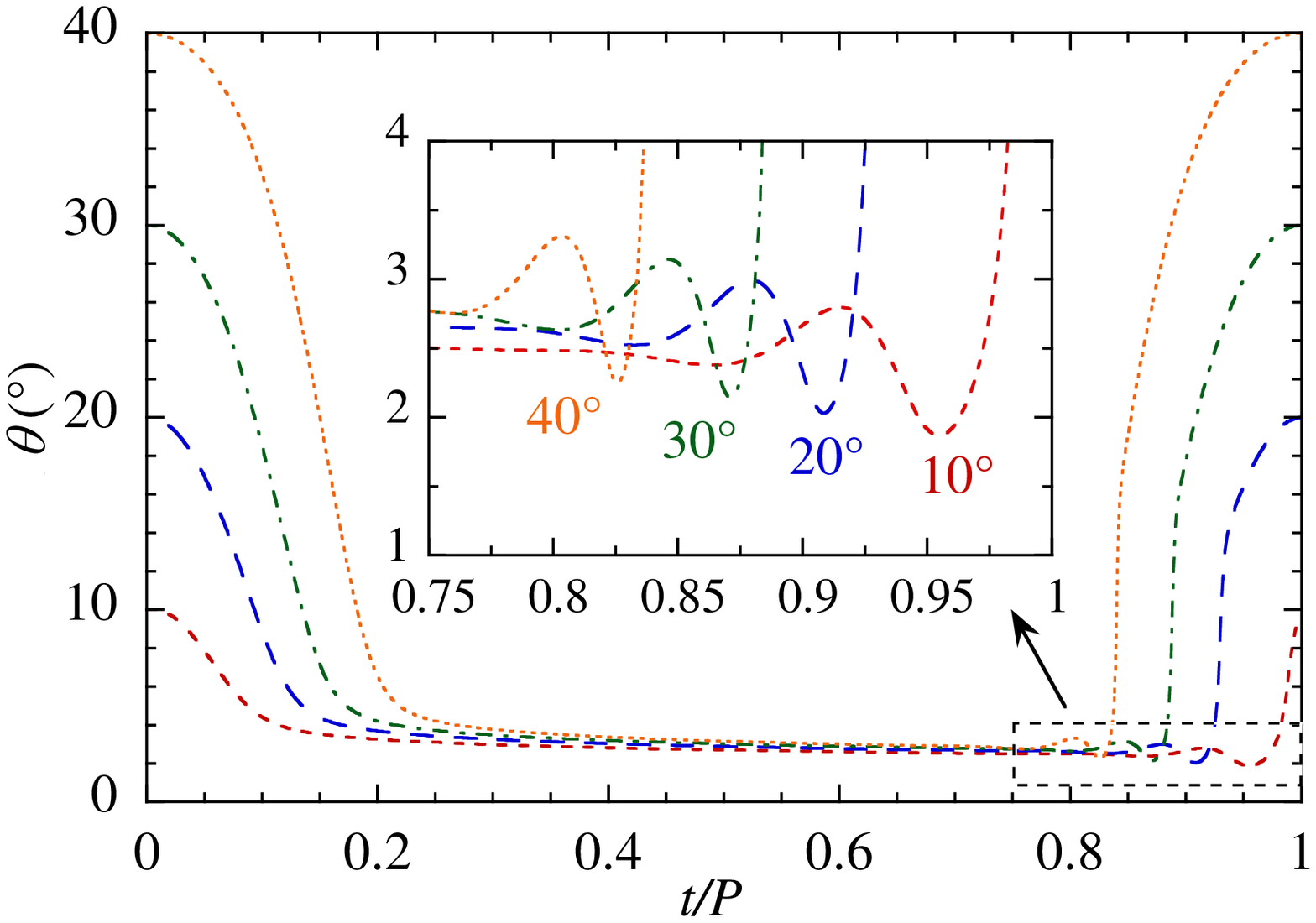}\label{fig:theta_app-vs-t}}\hfill
 \subfigure[ ]{\includegraphics[width=0.45\textwidth]{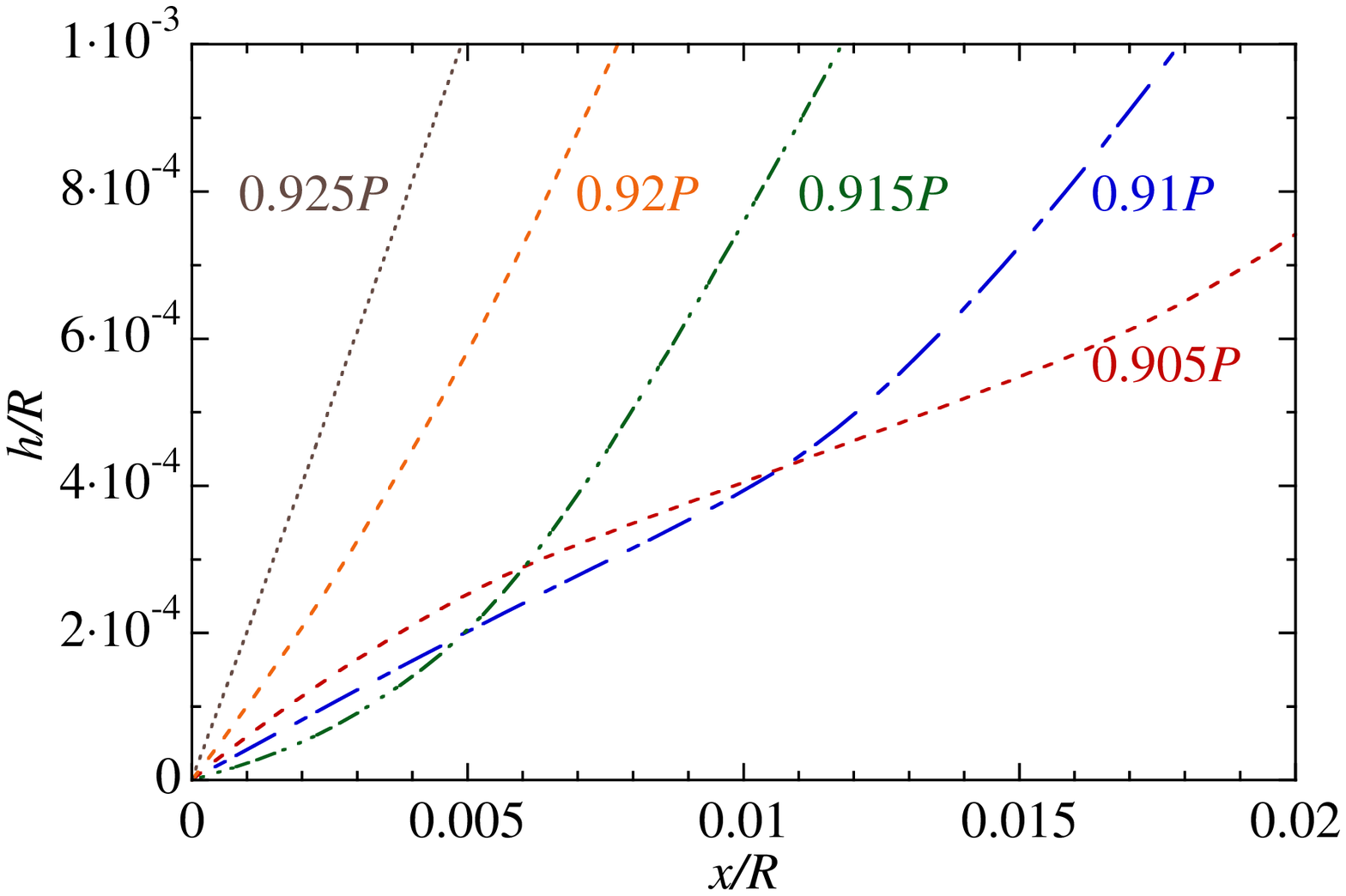}
 \label{hmin_approchingCL_P50}}
  \caption{Contact angle variation for $Ca_0=10^{-3}$, $\tilde P=50$. (a) Variation of contact angle during an oscillation for different values of $\theta_\mathrm{i}$. The inset shows enlarged undulating portions of the curves. (b) The interface shape variation near the contact line around $t=0.913P$ where $\theta_\mathrm{min}$ is attained for $\theta_\mathrm{i}=20^\circ$ (cf. the inset to Fig.~\ref{fig:theta_app-vs-t}).}
\end{figure}

A typical variation of $\theta$ during oscillation is plotted in Fig.~\ref{fig:theta_app-vs-t}. The initial contact angle $\theta_\mathrm{i}$ is the maximum contact angle achieved during the periodic motion.

In the beginning of a period, the capillary forces lead to the fast contact angle reduction until the meniscus recedes far enough so the curvature gradient reduces and the contact angle becomes nearly constant for a large part of a period. This nearly constant value is quite insensitive to both $\theta_\mathrm{i}$ and $P$. A small ridge of constant curvature forms near the contact line. This phenomenon is similar to the dewetting ridge but of much smaller magnitude because the contact line is pinned. A small ridge can be seen in Fig.~\ref{P=150_h_vs_x} in the contact line region for the curve corresponding to $\theta_\mathrm{i}=40^\circ$.

\begin{figure}
  \centering
  \subfigure[ ]
  {\includegraphics[width=0.47\textwidth]{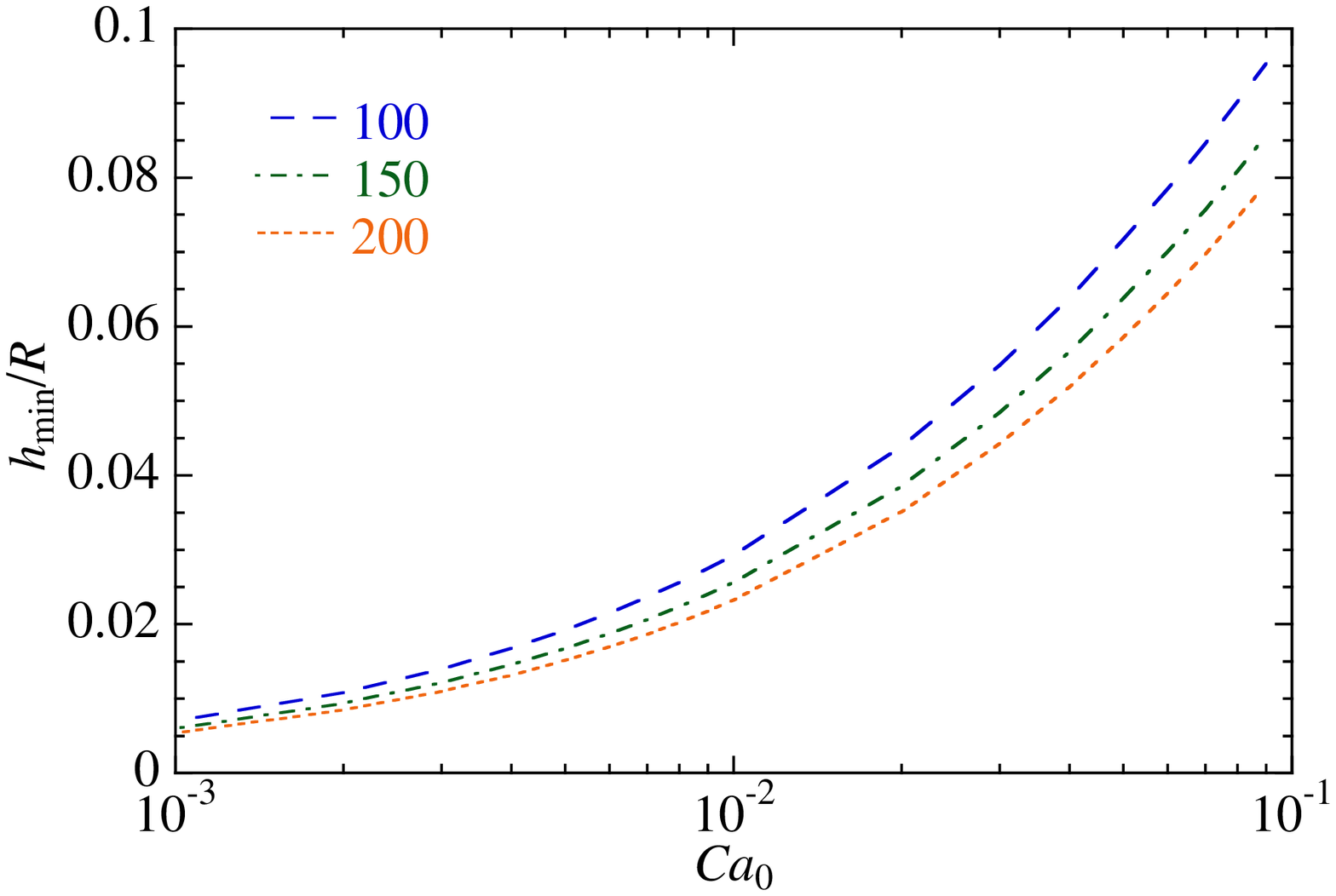}\label{h_min_vs_Ca}}\hfill
  \subfigure[ ]
  {\includegraphics[width=0.45\textwidth]{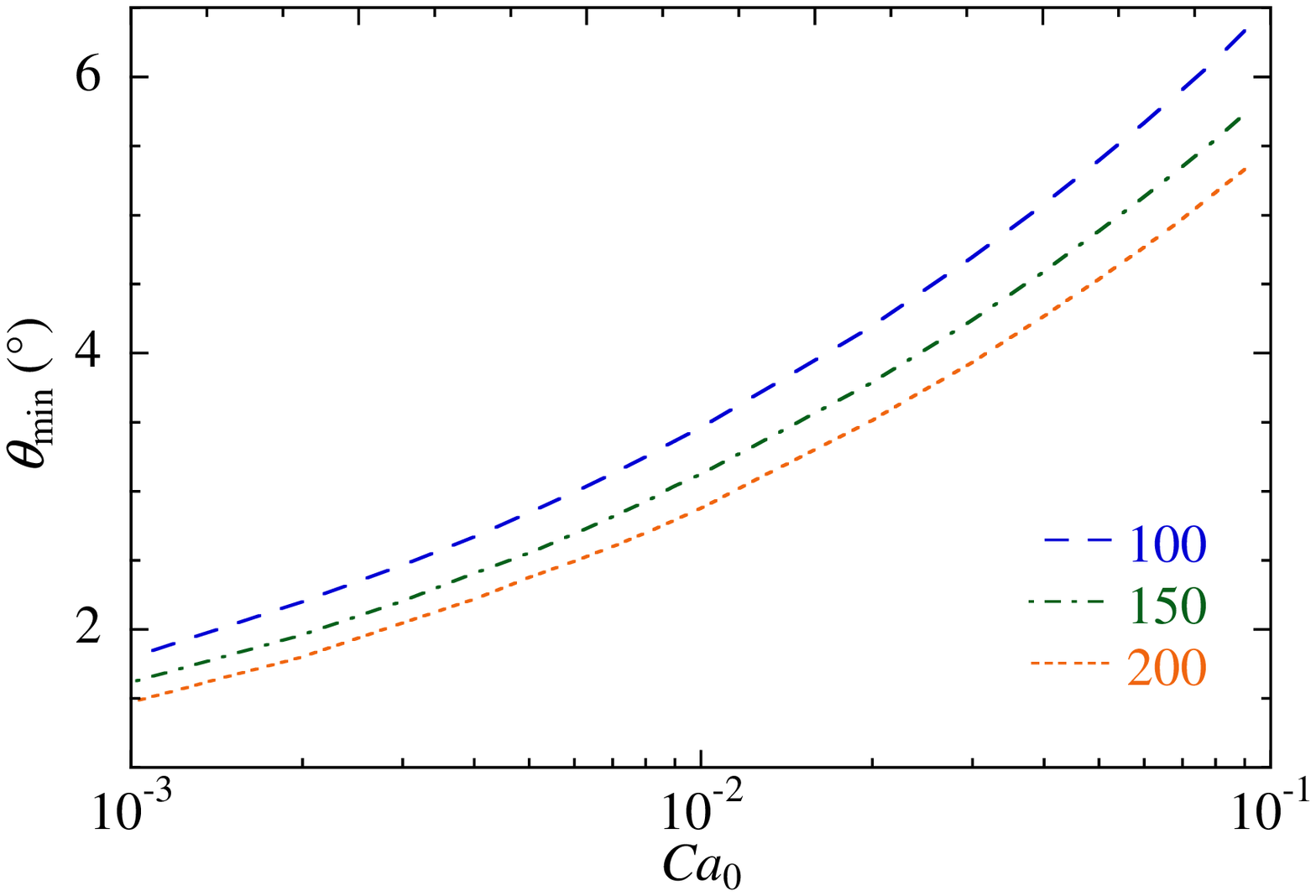}\label{CA_min_vs_Ca}}
  \caption{Variations of $h_\mathrm{min}$ and $\theta_\mathrm{min}$ with $Ca_0$ for $\theta_\mathrm{i}=20^\circ$ and different $\tilde P$; (a) $h_\mathrm{min}$ variation with $Ca_0$ and (b) $\theta_\mathrm{min}$ variation with $Ca_0$.}\label{min_vs_Ca}
\end{figure}

The ridge width slowly grows so $\theta$ slowly decreases until the ripples in the near-meniscus region approach the contact line during the backward stroke (Fig.~\ref{hmin_approchingCL_P50}) at the end of a period. This causes the contact angle oscillations, during which its minimal value $\theta_\mathrm{min}$ is attained
\begin{equation}\label{eq:theta_min}
  \theta(t) \geq \theta_\mathrm{min}.
\end{equation}
It depends quite weakly on $\theta_\mathrm{i}$ (Fig.~\ref{fig:theta_app-vs-t}). The variation of $\theta_\mathrm{min}$ with the system parameters follows the variation of $h_\mathrm{min}$ (which is a minimum of $h(x, 0.5P)$ observed near the meniscus). This is illustrated in  Fig.~\ref{min_vs_Ca}, where the variations of $h_\mathrm{min}$ and $\theta_\mathrm{min}$ with the system parameters are compared. Only the variations with $\tilde P$ and $Ca_0$ are considered (as mentioned above, the dependence on $\theta_\mathrm{i}$ is quite weak). Evidently, the variations of $h_\mathrm{min}$ and $\theta_\mathrm{min}$ with the system parameters are similar, which shows their intrinsic link. At $Ca_0\to 0$, the dependence on $Ca_0$ is weak, but becomes stronger at large $Ca_0$. This minimal value of the contact angle is of importance (cf. sec.~\ref{oscevapsec} below). Since the motion is periodic, the contact angle $\theta_\mathrm{i}$ is attained at $t=P$.


\subsection{Meniscus curvature during oscillation}\label{CurvResSec}

\begin{figure}
  \centering
  \subfigure[ ]{\includegraphics[width=0.45\textwidth]{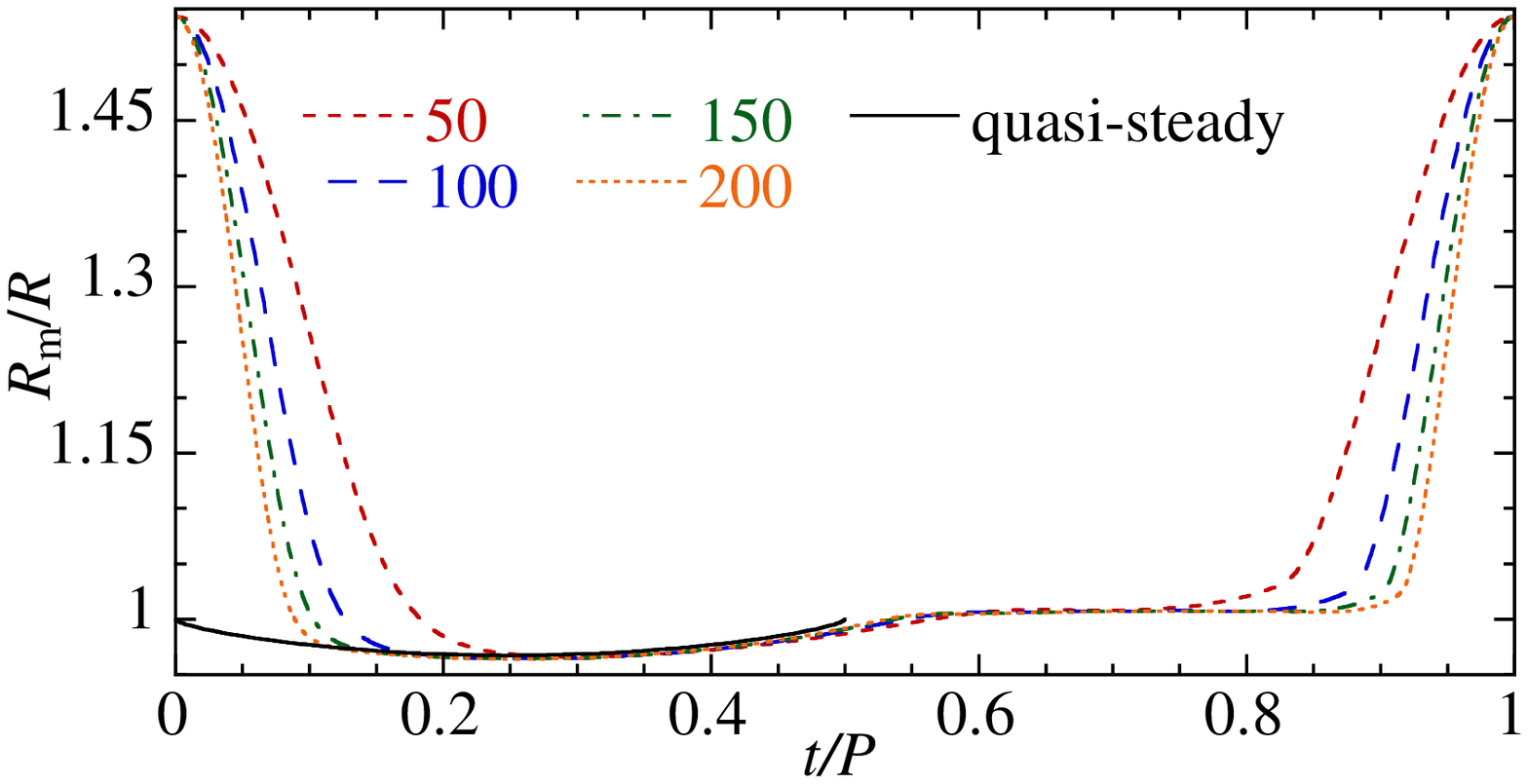}\label{Rm-vs-t_CA=40}}\hspace*{5mm}
  \subfigure[ ]{\includegraphics[width=0.45\textwidth]{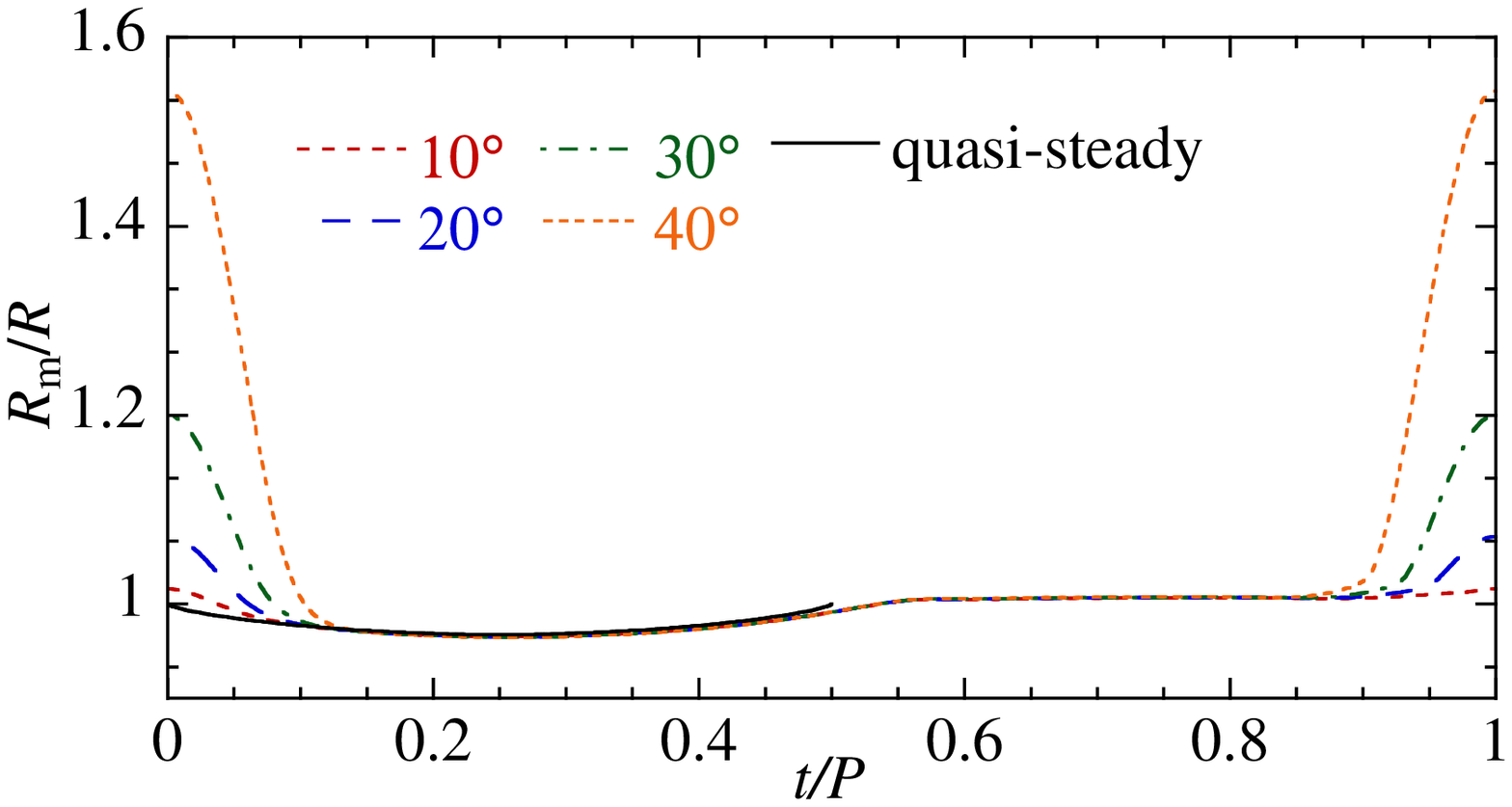}\label{Rm-vs-t_P=150}}
  \subfigure[ ]{\includegraphics[width=0.45\textwidth]{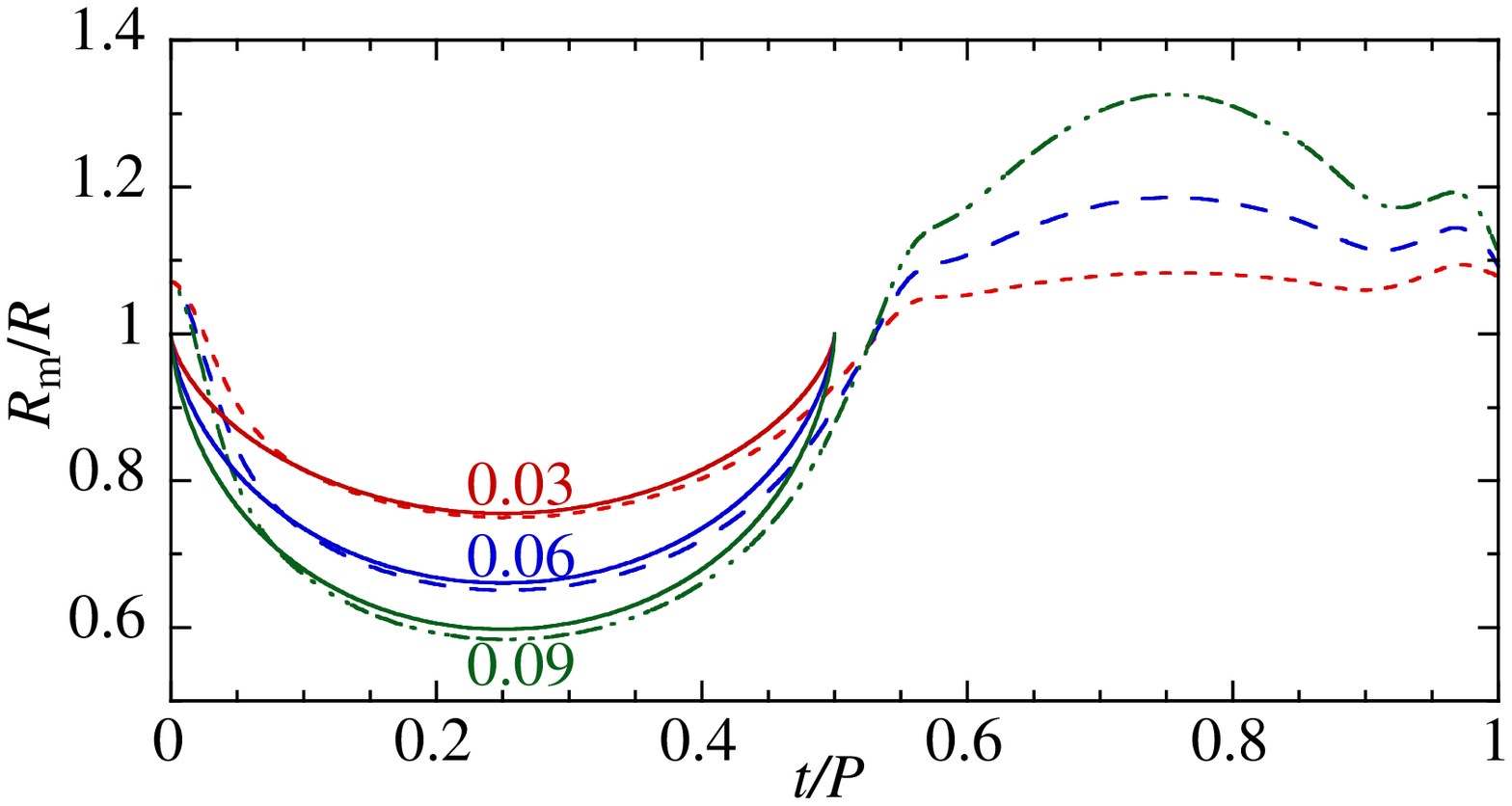}\label{Rm_vs_t_various_Ca}}
  \caption{Time evolution of the radius of meniscus curvature during an oscillation. The quasi-steady evolution of $R_\mathrm{m, q}$ at receding ($t\leq 0.5P$) is shown for comparison: (a) $R_\mathrm{m}$ evolution for different $\tilde P$;  $\theta_\mathrm{i}=40^\circ$ and $Ca_0=10^{-3}$ are fixed, (b) $R_\mathrm{m}$ evolution for different $\theta_\mathrm{i}$; $\tilde P =150$ and $Ca_0=10^{-3}$ are fixed and (c) $R_\mathrm{m}$ evolution (dashed curves) for different $Ca_0$;  $\tilde P =150$ and $\theta_\mathrm{i}=20^\circ$ are fixed. The solid curves of the respective colour show $R_\mathrm{m, q}(t)$.}\label{Rm-vs-t}
\end{figure}

The meniscus curvature (Figs.~\ref{Rm-vs-t}) changes periodically during oscillations. The value of $R_\mathrm{m}$ can be compared to the quasi-steady value $R_\mathrm{m, q}$ given by Eq.~\eqref{eq:crv-receding} where $\tilde h_\mathrm{s}=2.5$ and $Ca_r$ are calculated with the instantaneous meniscus velocity $U = U_0\sin(2\pi \tilde t/\tilde P)$ during the liquid receding ($t\leq 0.5P$).

For $t=0$, $R_\mathrm{m}$ is defined with Eq.~\eqref{eq:curvi}, which differs from the quasi-steady value $R_\mathrm{m, q}=R$ for $U=0$. This difference occurs because of the contact line presence. In its absence (pre-wetted tube, the situation equivalent to the limit $\theta_\mathrm{i}\to 0$ in our model), the initial radius would be close to $R$ because the wetting film is much thinner than the film considered here.

Within the time scale $\sim 5t_\mathrm{rel}$ (see Appendix~\ref{sec:relax}), $R_\mathrm{m}$ relaxes to the quasi-steady value $R_\mathrm{m, q}$. The curvature $R_\mathrm{m}$ remains close to $R_\mathrm{m, q}$ until the deceleration that occurs near the rightmost meniscus position (at $t=0.5P$, where $U=0$). However, the shape relaxation causes a delay, so the inequality $R_\mathrm{m}<R_\mathrm{m, q}(U=0)=R$ always holds at the point $t=0.5P$. During the backstroke, $R_\mathrm{m}$ varies, finally attaining the initial value \eqref{eq:curvi} that depends only on $\theta_\mathrm{i}$ (Fig.~\ref{Rm-vs-t_P=150}). Evidently, the amplitude of $R_\mathrm{m}$ oscillation grows with $Ca_0$ (Fig.~\ref{Rm_vs_t_various_Ca}), as foreseen by Eq.~\eqref{eq:crv-receding}. One also mentions the non-monotonic $R_\mathrm{m}$ variation during the backstroke with a local minimum around $t\simeq 0.8-0.9P$ (Fig.~\ref{Rm_vs_t_various_Ca}). This minimum appears when the trough of film ripples approaches the contact line close enough and is thus correlated with the contact angle minimum.

\subsection{Quasi-steady approach and average film thickness}\label{QuasiSec}

One can see that the film is thickest in its centre (Figs.~\ref{fig:film-CAs}), which correlates with the maximum of the meniscus velocity. It is thus interesting to compare the film thickness with its quasi-steady value. Within the quasi-steady approach, the term $\partial h/\partial t$ is neglected and the quasi-steady thickness $h_\mathrm{q}(x)$ can be defined as corresponding to the meniscus receding velocity as if it was constant at each time moment \citep{LauraATE17,Youn18}. In sec.~\ref{CurvResSec} it has been shown that a simple quasi-steady approach predicts well the meniscus curvature for $0<t<0.5P$. It is interesting to see if it is efficient in predicting the film profile. In this section, only the profile at $t=0.5P$ is considered, i.e. that with the longest film.
%

A difficulty appears because the film thickness $h(x,0.5P)$ depends on $x$. It is clear that $h$ depends on the velocity that the meniscus had at film deposition, but at which time moment? The most obvious first option is a moment $t$ when the meniscus centre was at the point $x$ (i.e. $x_\mathrm{m}(t)=x$). A more sophisticated option is a moment $t'>t$ such that
\begin{equation}\label{DX}
x_\mathrm{m}(t')=x+\Delta x
\end{equation}
with $\Delta x>0$. In the previous approaches \citep{LauraATE17,Youn18}, the first option was used. By using Eq.~\eqref{xm} one easily finds that this assumption defines $U_\mathrm{q}(x)$ for $x\in (x_\mathrm{i},x_\mathrm{i}+2A)$. An obvious contradiction occurs at $x=x_\mathrm{i}+2A$ where $h_\mathrm{q}(x_\mathrm{i}+2A)=0$ because $U_\mathrm{q}(x_\mathrm{i}+2A)=0$, but the actual film thickness (or rather, the interface height) is $R$. Therefore, a more realistic $\Delta x>0$ should be defined. It is reasonable to choose $\Delta x$ to be the meniscus radius, i.e. $R_\mathrm{m}$. However, it varies with time. We propose to use $\Delta x = \langle R_\mathrm{m} \rangle$, the average value of $ R_\mathrm{m}(t)$ (defined using Eq.~\ref{eq:crv-receding}) over the first half-period.

For the harmonic meniscus oscillation, one derives from Eqs.~(\ref{xm},\ref{DX}):
\begin{equation}\label{Uq}
U_\mathrm{q}(x)=U_0\left[\frac{ x +\Delta x -x_\mathrm{i}}{ A}\left(2-\frac{ x + \Delta x-x_\mathrm{i}}{A}\right)\right]^{1/2}.
\end{equation}
This velocity can now be used to calculate $Ca_r$ in Eq.~\eqref{eq:hr}, thus resulting in the quasi-steady thickness $h_\mathrm{q}(x)$. The $\tilde h_\mathrm{q}(\tilde x')$ profiles are shown as solid curves in Figs.~\ref{fig:film-CAs}. Note that they are independent of $\theta_\mathrm{i}$ (cf. Eq.~\ref{xprime}), so there is a unique curve in each figure. One can also see the necessity of the $\Delta x$ introduction: if it were not included, the $\tilde h_\mathrm{q}(\tilde x')$ curve would be shifted with respect to $\tilde h(\tilde x')$ curves in Figs.~\ref{fig:film-CAs}. The agreement between the actual and quasi-steady film profiles is good in the central part of the film, which shows that the $\Delta x$ choice is acceptable.

At the beginning of a period, due to the presence of the contact line, $R_\mathrm{m}$ is always larger than $R_\mathrm{m, q}$, which leads to a thinner film: $ h <  h_\mathrm{q}$, cf. Figs.~\ref{fig:film-CAs}. This difference becomes relatively less important as $P$ increases (compare Figs.~\ref{P=50_h_vs_x} and \ref{P=150_h_vs_x}).

\begin{figure}
  \centering
  \subfigure[ ]{\includegraphics[width=0.45\textwidth]{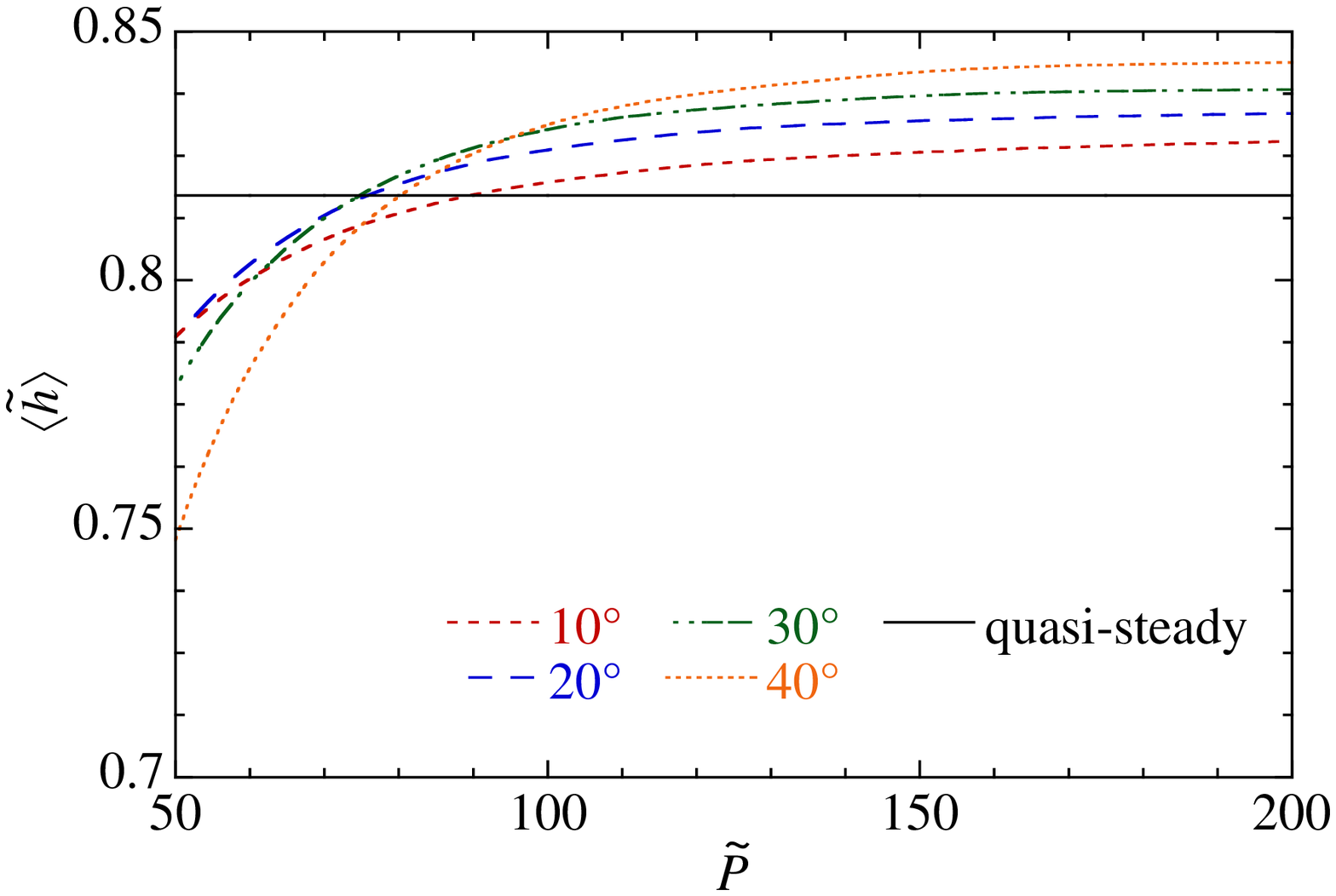}\label{avhtheta}}\hspace*{5mm}
  \subfigure[ ]{\includegraphics[width=0.45\textwidth]{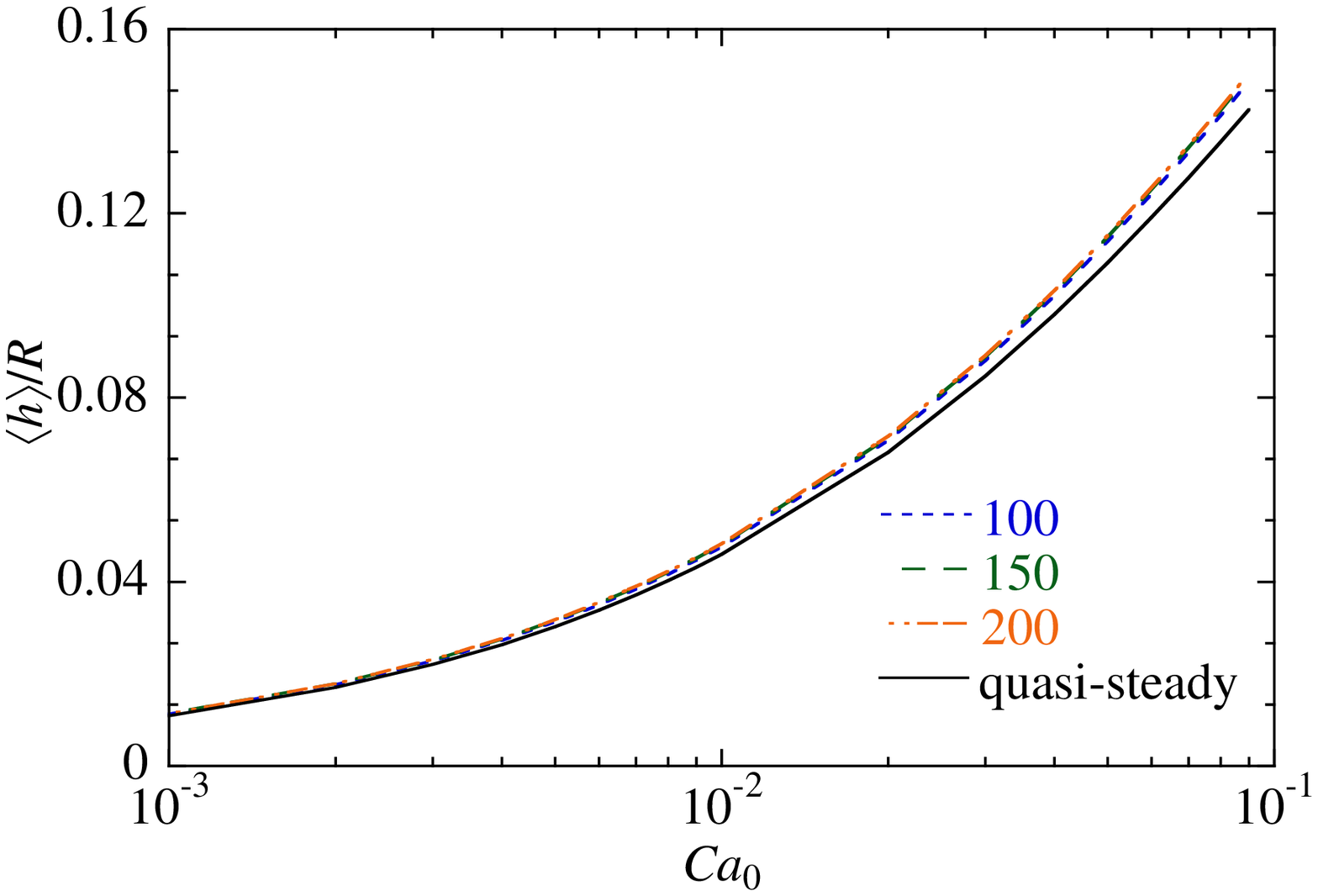}\label{fig:avhLargeCa}}
  \caption{Average thickness of the liquid film at $t=0.5P$; $\langle h_\mathrm{q}\rangle$ is shown for comparison. (A) Value of $\langle\tilde h\rangle$ as a function of oscillation period for different $\theta_\mathrm{i}$ and for $Ca_0=10^{-3}$ and (b) value of $\langle h\rangle$ as a function of $Ca_0$ for different $\tilde P$ and for $\theta_\mathrm{i}=20^\circ$.}\label{fig:averaged-h}
\end{figure}
To characterise the film at oscillation conditions (e.g. to estimate the film evaporation rate), it is important to know the average film thickness $\langle h\rangle$. It is defined as a spatial average over the interval between the contact line and $x_\mathrm{min}$ (abscissa of the point where $h_\mathrm{min}$ is attained),
\begin{equation}\label{hav}
 \langle h\rangle=\frac{1}{ x_\mathrm{min}}\int_0^{ x_\mathrm{min}} h( x, t=0.5 P)\mathrm{d} x.
\end{equation}

The dependence of $\langle h\rangle$ on different parameters can be seen in Figs.~\ref{fig:averaged-h}. It is an increasing function of $\tilde P$ that saturates for $\tilde P\to\infty$. It can be compared to the quasi-steady averaged thickness
\begin{equation}\label{hqav}
\langle h_\mathrm{q}\rangle=\frac{1}{2 A}\int_{x_\mathrm{i} - \Delta x}^{2 A+x_\mathrm{i} - \Delta x} h_\mathrm{q}( x)\mathrm{d} x
\end{equation}
which is independent of both $A$ and $\theta_\mathrm{i}$ as follows from Eq.~\eqref{Uq}. It can be easily calculated without doing any complicated simulations. For small $\tilde P$ (i.e. for small $\tilde A$), $\langle h\rangle<\langle h_\mathrm{q}\rangle$, mainly because of the contact line vicinity where $h(x)< h_\mathrm{q}(x)$, cf. Figs.~\ref{fig:film-CAs}. It is not surprising that the saturation value of $\langle h\rangle$ increases with $\theta_\mathrm{i}$ (just because of the thicker film near the contact line); however, the increase is weak (Fig.~\ref{avhtheta}).

In Fig.~\ref{fig:avhLargeCa}, one can see the dependence of $\langle h\rangle$ on $Ca_0$. One can see that the quasi-steady average $\langle h\rangle _\mathrm{q}$ Eq.~\eqref{hqav} gives a globally satisfactory approximation of $\langle h\rangle$. The increase with $Ca_0$ is mainly due to the $Ca_0^{2/3}$ factor in Eq.~\eqref{eq:hinf}.

\subsection{Film shape comparison with experiment}\label{expqual}

In the experiments of \citet{Lips10}, a capillary tube contains a short liquid plug of pentane in contact with its own vapour at both ends of the tube. One end is connected to a reservoir at constant pressure. The pressure variation at the other end forces the oscillating motion of a liquid plug under isothermal conditions. Such a mode of oscillation leads to the oscillation amplitude increasing in time, which was understood some years later \citep{Mamba18}. While the amplitude is indeed slightly increasing, the motion is nearly periodic, so the comparison can still be done.

\begin{figure}
  \centering
  \includegraphics[width=0.5\textwidth,clip]{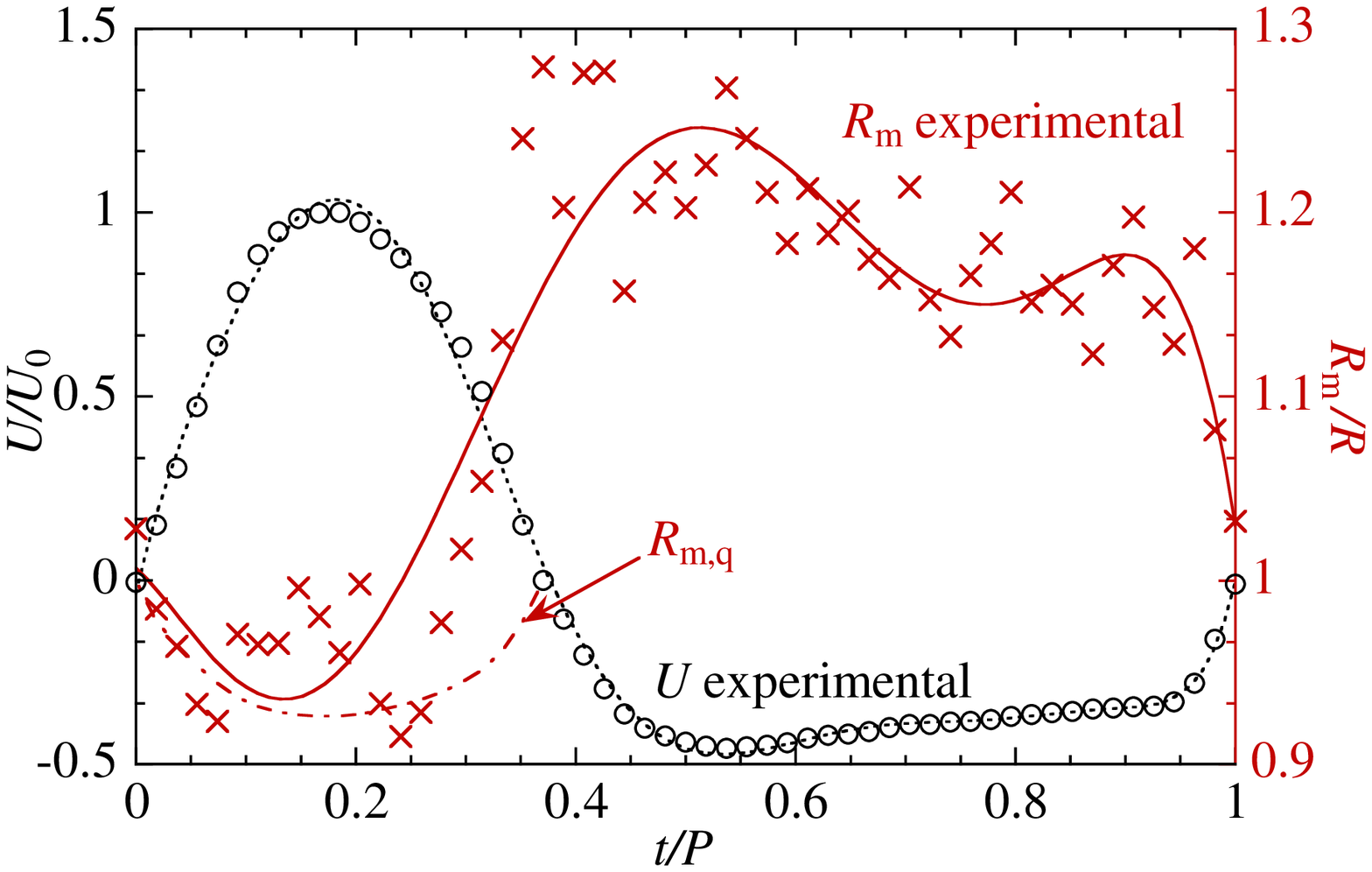}
  \caption{Liquid plug velocity and left meniscus radius variations during one period of the \citet{Lips10} experiment (characters) and their polynomial fits (lines) used for the calculation. The experimental data have been made available to us by S. Lips. For comparison, the quasi-steady $R_\mathrm{m,q}$ variation given by Eq.~\eqref{eq:crv-receding} at receding ($t\leq 0.5P$) is also shown.}\label{fig:lips10_U_R_vs_t}
\end{figure}

In their experiments, the plug motion is recorded with a high-resolution camera. Both the meniscus velocity and the curvature radius are found from image analysis. The Weber number $We_0= 2R\rho U_0^2/\sigma$ is larger than unity (table~\ref{tab:lips10_values13}). The Reynolds number $\Rey_0=2R\rho U_0/\mu$ is quite high too and the impact of inertia on the shape of the central meniscus part must be taken into consideration. Indeed, the quasi-steady $R_\mathrm{m}(t)$ evolution and the experimental measurements of \citet{Lips10} differ (Fig.~\ref{fig:lips10_U_R_vs_t}). One mentions that the measured $R_\mathrm{m}\simeq R$ at $t=0, P$, which indicates the complete wetting case. Note the $R_\mathrm{m}$ local minimum around $t\simeq 0.8P$. A similar minimum appears in the simulation, see Fig.~\ref{Rm_vs_t_various_Ca} and the associated discussion.

In spite of high $We_0$ and $\Rey_0$ values mentioned above, the thin film can still be considered as controlled by the viscosity only. In the simulation, instead of using the $R_\mathrm{m}$ calculation of sec.~\ref{CurvSec}, the experimental plug velocity and the radius variation shown in Fig.~\ref{fig:lips10_U_R_vs_t} are used. Under these conditions, the film ripples in the transition region close to the meniscus can be compared to the calculations.

\begin{table}
\centering
\begin{tabular*}{0.6\textwidth}{@{\extracolsep{\fill}}lcl}

parameter         & notation      & value \\[3pt]
tube inner radius &    $R$        & 1.2 mm        \\
density           &$\rho$         & 625.7 kg/m$^3$        \\
surface tension   &$\sigma$                   & 0.0152 N/m        \\
shear viscosity &$\mu$              & $2.37\cdot 10^{-4}$ Pa$\cdot$s        \\
reference velocity (velocity amplitude)&$U_0$              & 0.24 m/s        \\
oscillation frequency           &                      & 3.7Hz         \\
dimensionless period           &   $\tilde P$           & 260.1         \\
capillary number  &$Ca_0$       & $3.74\cdot 10^{-3}$\\
Reynolds number   &$\Rey_0$           &1521\\
Weber number      &$We_0$           &5.66\\

\end{tabular*}
\caption{Fluid properties at the experimental conditions (1bar, $20^\circ$C) of \citet{Lips10};  and key dimensionless numbers.}\label{tab:lips10_values13}
\end{table}

\begin{figure}
  \centering
  \includegraphics[width=0.9\textwidth,clip]{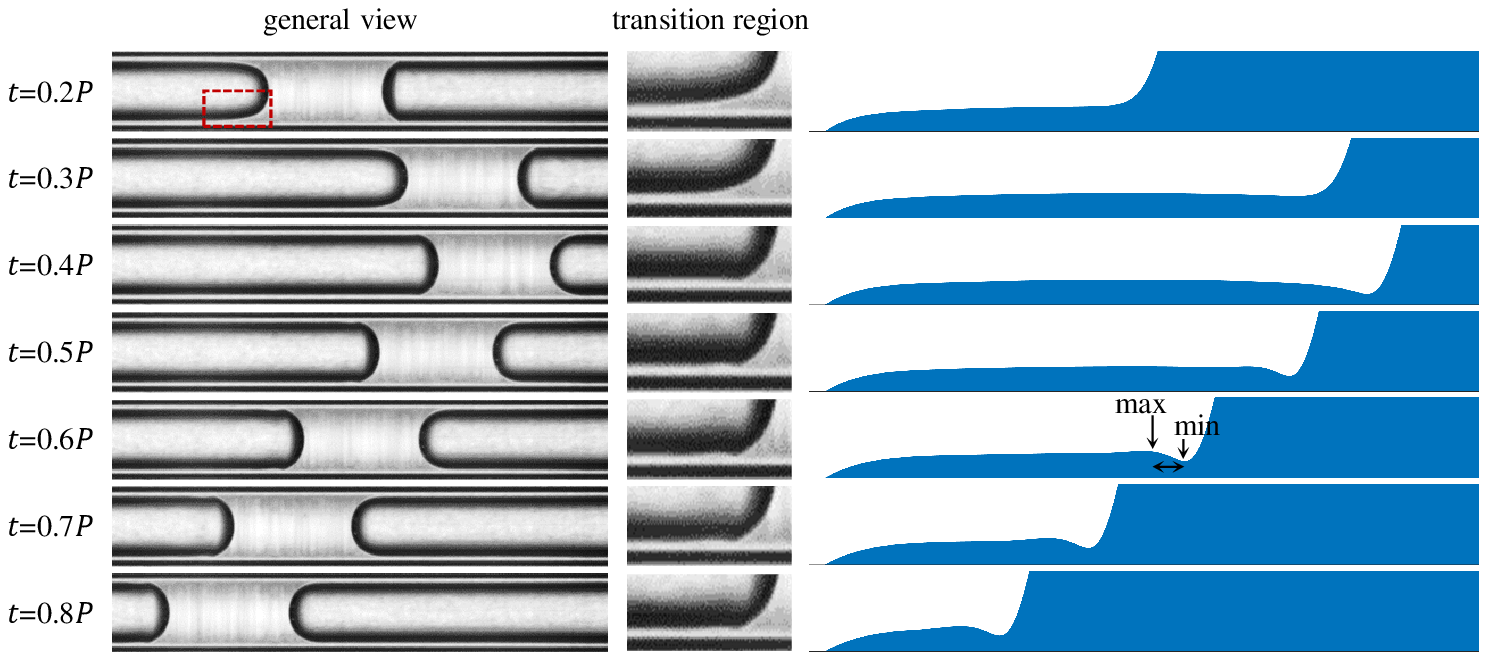}
  \caption{Liquid film shape during meniscus oscillation; two left columns: experimental results by \citet{Lips10}; the right column: numerical results (see the supplementary movie).}\label{fig:lips_simu_comparison}
\end{figure}
Figure~\ref{fig:lips_simu_comparison} presents several snapshots of plug oscillation. The left column shows the original images of \citet{Lips10}. The liquid film shape in the transition region between the film and the meniscus is enlarged in the middle column. The numerical results are shown in the right column. One can see that the wavy appearance of the interface is truthfully captured by the numerical calculation.

Unfortunately, the quantitative comparison of the film thickness (i.e. the vertical coordinate) is hardly possible since the refraction by the glass capillary is not corrected and the spatial resolution is not sufficient to distinguish the contact line.  However, one can compare the axial lengths. The size of one pixel in mm can be obtained from the known outer tube diameter (4~mm) that is visible in the original images. One can compare the axial distance between the local maximum and the local minimum (Fig.~\ref{fig:lips_simu_comparison}) of the film ripple. It can be measured for the images corresponding to $t\geq 0.5P$ where the ripple is visible. The distance is almost constant in time. From the experimental images, the axial distance is $0.50\pm0.02$~mm, while from the simulation, it is $0.51\pm0.01$~mm. Evidently, the agreement is excellent.

\subsection{Film thickness comparison with experiment}

The experiment of \citet{Youn18} was carried out under adiabatic conditions. They investigated the deposited film thickness of an oscillating meniscus in a cylindrical capillary tube. Two working fluids are selected for the comparison with the present numerical results: water and ethanol. Fluid properties and experimental parameters are summarised in table~\ref{tab:Youn18}.

In the tests, the capillary tube is partially filled so the syringe piston is in contact with liquid; there is a single meniscus in the tube. The piston is connected to a step motor that imposes the harmonic motion. The other tube end remains open. Initially, the meniscus remains stationary, then, following the piston, starts to oscillate with a constant frequency. Several sensors that measure the film thickness are installed along the tube, within the range of meniscus oscillation. The instantaneous velocity of the meniscus when it passes the sensors is recorded by the high speed camera. Because of the open tube, film evaporation occurs; however, it is much slower than the oscillation velocity so the film thickness variation during the oscillation period is negligibly small. Unfortunately, the film shape near the meniscus and near the contact line was not studied in their experiments and thus cannot be compared with our results.

\begin{figure}
  \centering
  \subfigure[]{\includegraphics[width=0.48\textwidth]{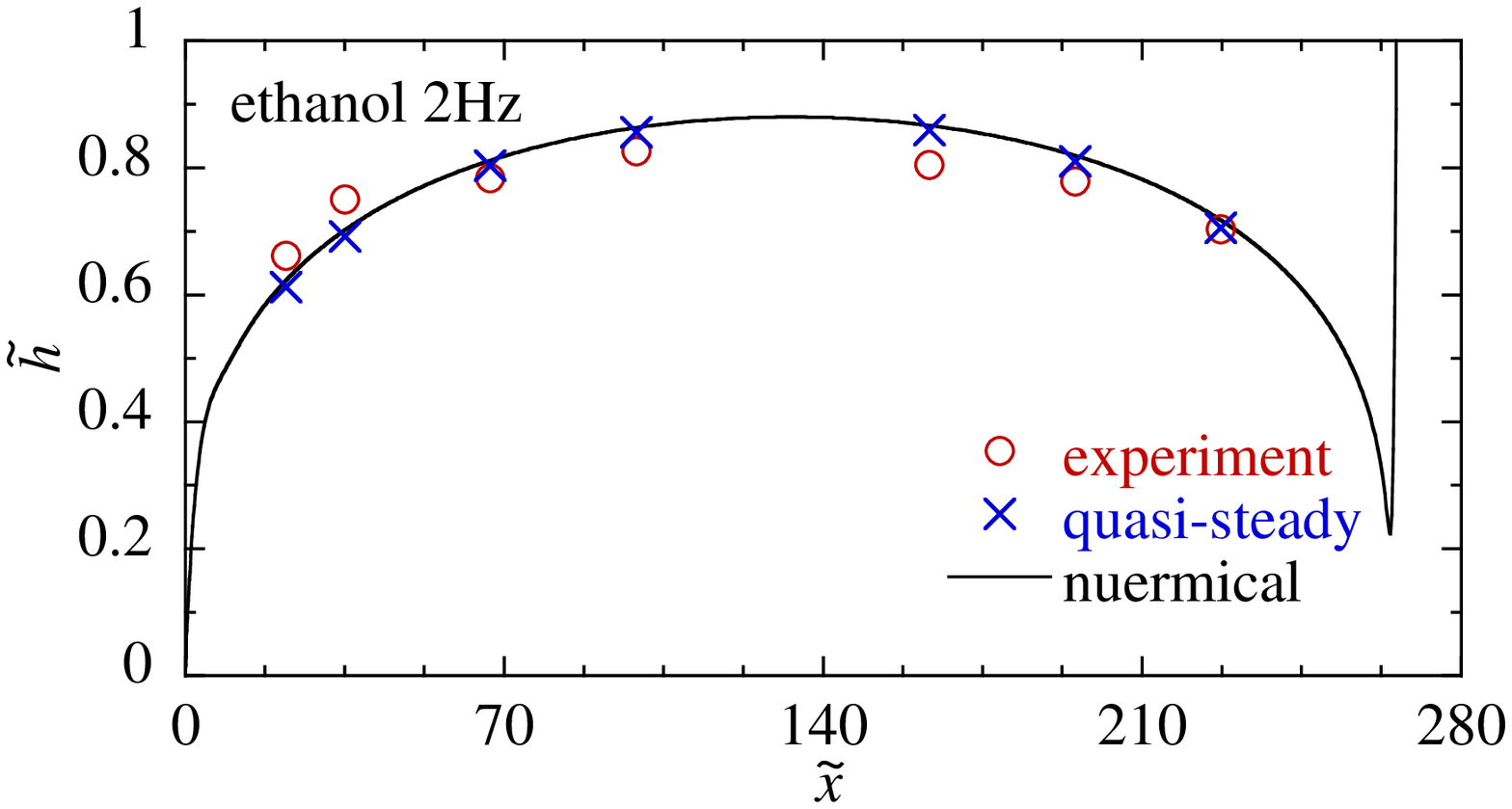}\label{ethanol2Hz}}\hspace*{5mm}
  \subfigure[]{\includegraphics[width=0.48\textwidth]{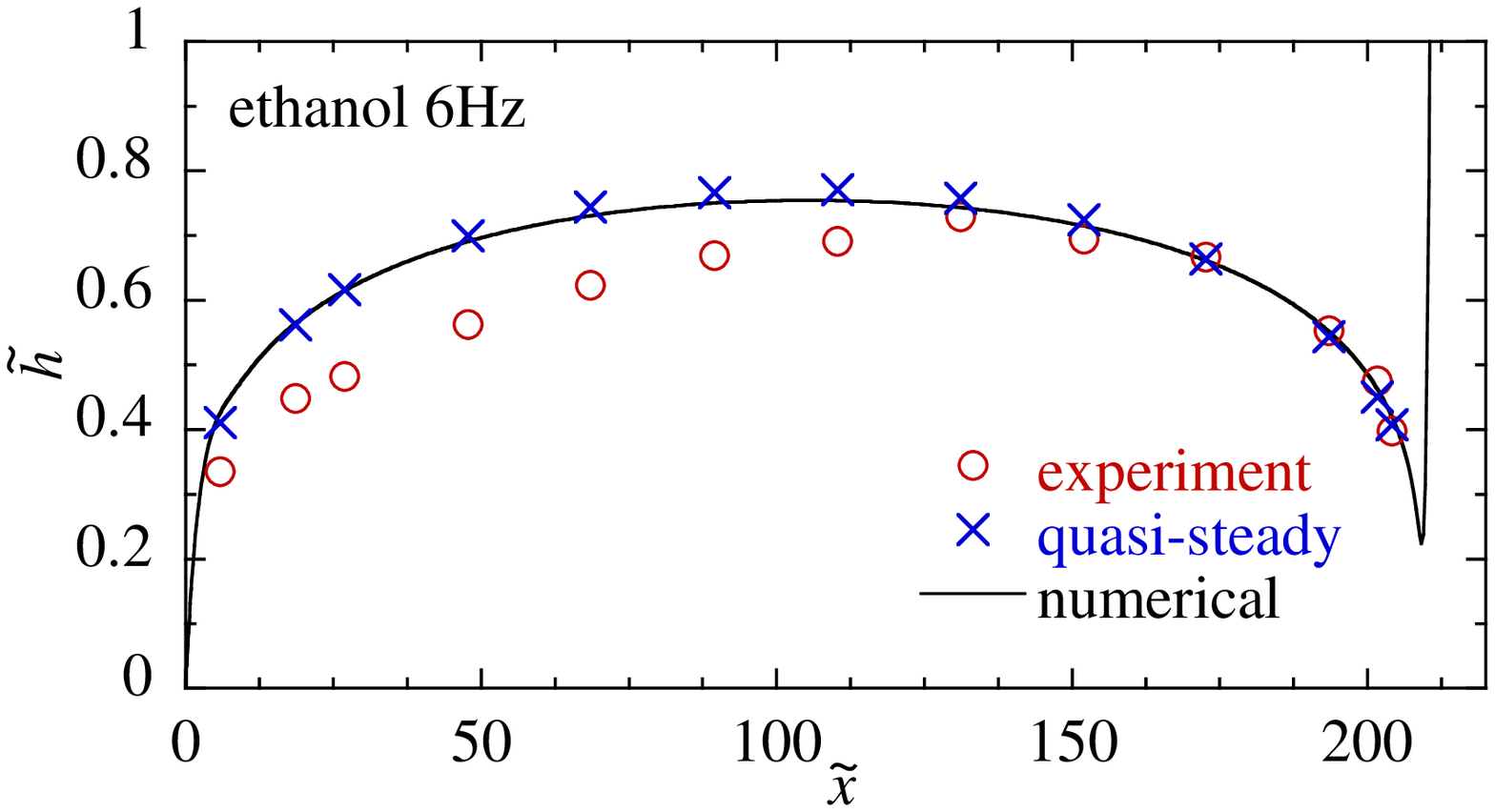}\label{ethanol6Hz}}
  \caption{Liquid film thickness: comparison of the experimental data of \citet{Youn18}, quasi-steady estimation with Eq.~\eqref{eq:hr} and numerical results for ethanol.}\label{Fig:ethanol}
\end{figure}

\begin{figure}
  \centering
  \subfigure[]{\includegraphics[width=0.48\textwidth]{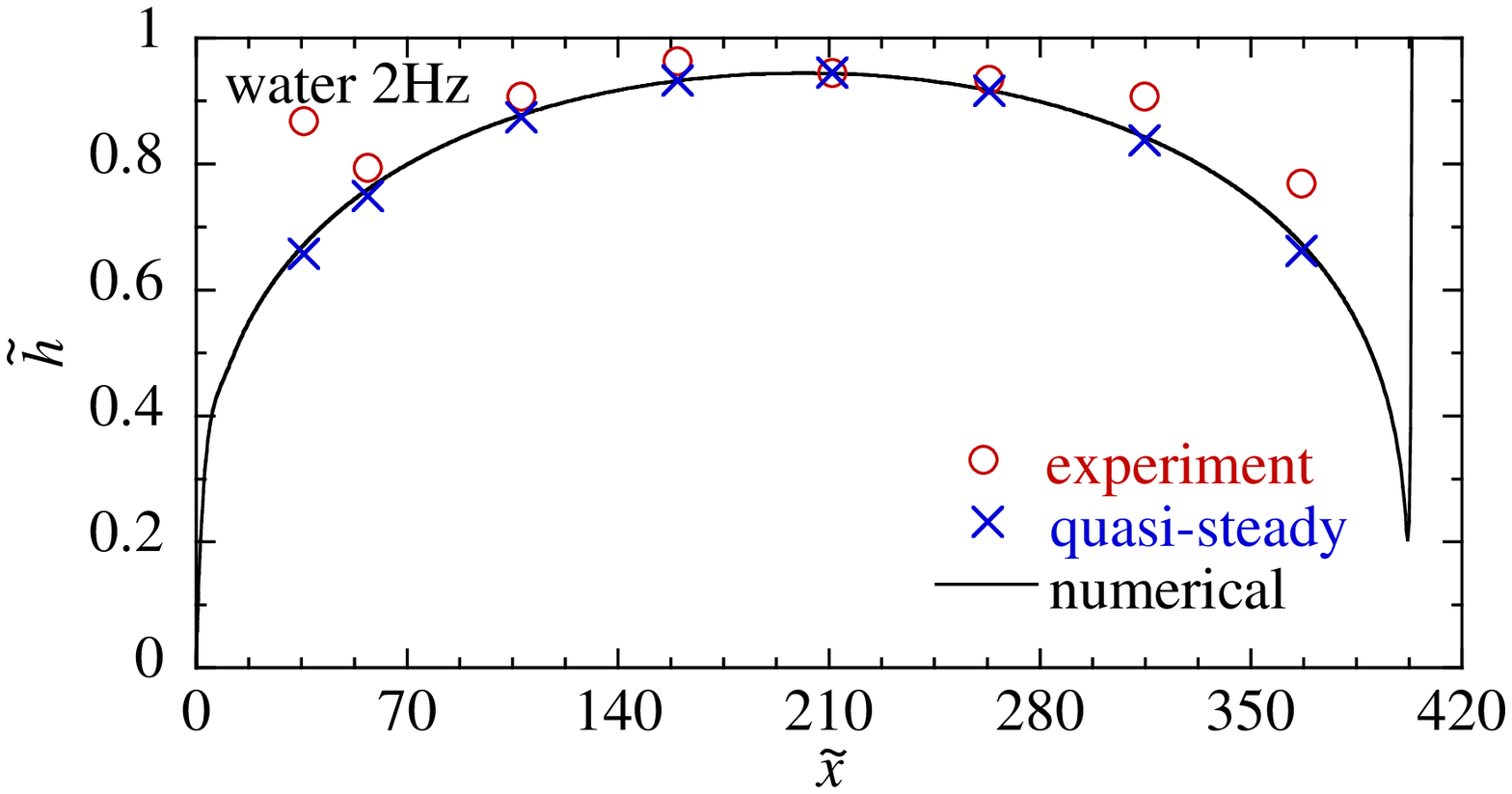}\label{water2Hz}}\hspace*{5mm}
  \subfigure[]{\includegraphics[width=0.48\textwidth]{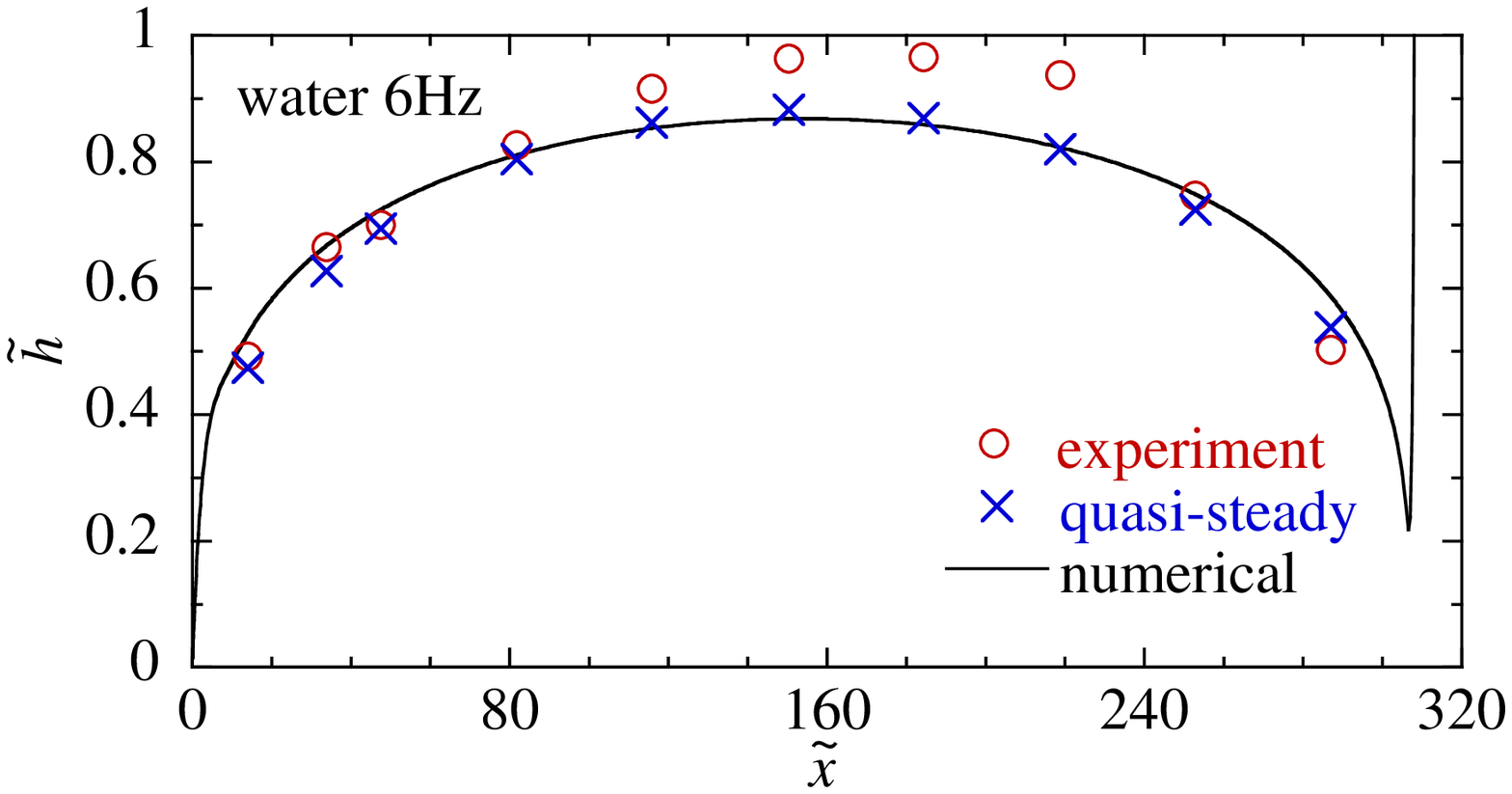}\label{water6Hz}}
  \caption{Liquid film thickness: comparison of the experimental data of \citet{Youn18}, quasi-steady estimation with Eq.~\eqref{eq:hr} and numerical results for water.}\label{Fig:water}
\end{figure}

The numerical calculation has been done as described in sec.~\ref{sec:osctheor} to get the film profiles at $t=0.5P$. The experimental film thickness and the numerical results are presented in Fig.~\ref{Fig:ethanol} for ethanol and in Fig.~\ref{Fig:water} for water. The experimental points are plotted with respect to the meniscus position known from the experimental data.

The experiment can also be compared with the quasi-steady film thickness calculated with Eq.~\eqref{eq:hr} (with $\tilde h_\mathrm{s}\simeq 2.5$) where the instantaneous value of the velocity \eqref{Uq} is used in $Ca_\mathrm{r}$.

\begin{table*}
\renewcommand\arraystretch{1.15}
  \centering
\begin{tabular*}{0.9\textwidth}{@{\extracolsep{\fill}}lccccc}

parameter         & notation             & \multicolumn{4}{c}{value}                               \\\hline
tube inner radius & $R$                  & \multicolumn{4}{c}{0.5 mm}                                 \\
fluid             &            {}        &\multicolumn{2}{c}{water} & \multicolumn{2}{c}{ethanol} \\ \cline{3-6}
surface tension   &    $\sigma$          & \multicolumn{2}{c}{$7.20\cdot 10^{-2} $ N/m} & \multicolumn{2}{c}{$2.23\cdot 10^{-2} $ N/m} \\
shear viscosity &     $\mu$            & \multicolumn{2}{c}{$8.88\cdot 10^{-4} $ Pa$\cdot$s}  & \multicolumn{2}{c}{$1.088\cdot 10^{-3} $ Pa$\cdot$s} \\
liquid density    &    $\rho$            & \multicolumn{2}{c}{997 kg/m$^3$}              & \multicolumn{2}{c}{785 kg/m$^3$}              \\
oscillation frequency &              & \multicolumn{1}{c}{2 Hz}  & \multicolumn{1}{c}{6 Hz} &\multicolumn{1}{c}{2 Hz} &\multicolumn{1}{c}{6 Hz}  \\ \cline{3-6}
oscillation amplitude &$A_0$             & 19.5 mm     & 22.5 mm     & 20.69 mm     & 25.2 mm      \\
$U$ amplitude in simulation   &$U_0=2\pi A/P$            & 0.245 m/s   & 0.848 m/s   & 0.26 m/s     & 0.950 m/s    \\
$U$ amplitude in experiments  &       & 0.254 m/s   & 1.025 m/s   & 0.285 m/s     & 1.150 m/s    \\
dimensionless period  &$\tilde P$   & 1267     & 966     & 834 &659      \\
capillary number    &$Ca_0$     & 0.0030      & 0.0105     & 0.0127       & 0.0463       \\
Reynolds number     &$\Rey_0$      & 275.07      & 952.09     & 187.59       & 685.43       \\
Weber number        &$We_0$        & 0.8312      &9.9576       & 2.3796       & 31.77        \\
\end{tabular*}
  \caption{Fluid properties at the experimental conditions (1bar, $25^\circ$C) of \citet{Youn18}. $A_0$ and the experimental $U$ amplitude are not explicitly given by them and are estimated from their graphs.}\label{tab:Youn18}
\end{table*}

Some experimental parameters are summarised in table \ref{tab:Youn18}. The numerical results are in excellent agreement with the quasi-steady data for 2 Hz, where both $We_0$ and $\Rey_0$ are moderate. This is not surprising as $\tilde P$ is large (see the discussion in sec.~\ref{QuasiSec}). Generally, there is a good agreement with the experimental data for the same reason. The discrepancy between the experimental and numerical results is larger for the 6 Hz case where both $We_0$ and $\Rey_0$ become large. Both dimensionless numbers are smaller for the ethanol than for the water, and the discrepancy is smaller too. One can conclude that the discrepancy is caused by the inertial effects that become important when $\Rey_0$ attains 500 and $We_0$ attains 10.

An additional discrepancy comes from the non-harmonicity of the meniscus velocity profile. As the experimental $U(t)$ curves are  unavailable, we use the harmonic law \eqref{xm} that corresponds to experimental oscillation period $P$ and amplitude $A$. Because of the film deposition, the experimental $U(t)$ deviates from the harmonic law. This can be observed from table \ref{tab:Youn18}. Indeed, the velocity amplitude $2\pi A/P$ calculated from the oscillation amplitude and the frequency differs substantially from the actual maximum velocity, which points out the non-harmonicity of the oscillations. The comparison would be improved if the experimental $U(t)$ were available to us.

\section{Combined effect of oscillation and evaporation}\label{oscevapsec}

In this section we show the implication of the above results for the case of heating conditions (i.e., a positive $\Delta T$). According to Eq.~\eqref{J}, evaporation occurs for $\Delta T>0$, and the mass flux $J(x)\propto h^{-1}(x)$. Instead of solving Eq.~\eqref{eq:GE-kelvin} for this case (which is out of the scope of the present article), we apply here the multi-scale reasoning introduced in sec. \ref{RelaxSec}. We consider below a case of a small $\Delta T$ so the evaporation in the film and meniscus regions can be neglected during an oscillation period so the macroscopic results of sec. \ref{sec:osctheor} still apply. However, because of the above singularity, the effect is not negligible in the microregion (inner region) and leads to a difference between the apparent contact angle $\theta$ and the microscopic contact angle $\theta_\mathrm{micro}$, as defined in sec.~\ref{RelaxSec}. Their relationship can be expressed as
\begin{equation}\label{eq:theta_T_theta_mic}
  \theta = \theta(\Delta T, \theta_\mathrm{micro}),
\end{equation}
cf. Appendix~\ref{sec:app}.


When a solution of the full problem exists, the micro- and macroregions can be matched for a given $\Delta T$. They are connected through the formula~\eqref{eq:theta_T_theta_mic}, which means that the oscillatory film shape eventually enforces the value of $\theta_\mathrm{micro}$.

In the sense of the multi-scale reasoning, $\theta$ must simultaneously satisfy the conditions:
\begin{enumerate}
  \item Similarly to the case where $\Delta T =0$, the dynamic film shape imposes the $\theta$ value because the contact line is fixed. This implies that the inequality \eqref{eq:theta_min} should hold for $\Delta T\neq 0$ too.
  \item On the other hand, $\theta$ is defined by Eq.~\eqref{eq:theta_T_theta_mic}. As shown in Appendix~\ref{sec:app}, it is bounded from below: $\theta \geq \theta_\mathrm{evp-min} (\Delta T)$.
\end{enumerate}

\begin{figure}
  \centering
  \includegraphics[width=0.5\textwidth]{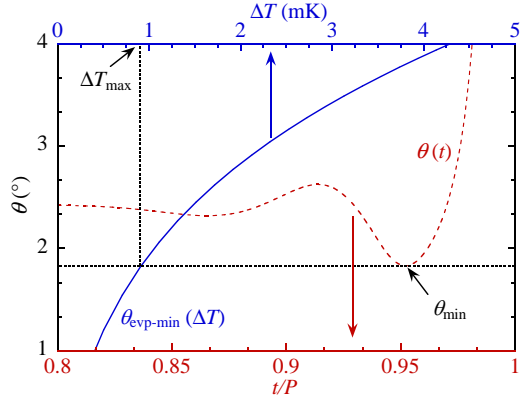}
  \caption{Apparent contact angle $\theta$ calculated for $\theta_\mathrm{micro}=0$ as a function of superheating $\Delta T$ (example of pentane at 1bar, same as in Fig.~\ref{fig:ThetaApp-dT}) plotted together with the apparent contact angle variation during meniscus oscillation. The latter is a curve for initial value $10^\circ$, the same as at the inset of Fig.~\ref{fig:theta_app-vs-t}.}
  \label{fig:theta_app_c-vs-deltaT-and-theta_app-vs-t}
\end{figure}

Condition (ii) means that $\theta (t)$ remains larger than $\theta_\mathrm{evp-min}(\Delta T)$ throughout oscillation. From the inequality~\eqref{eq:theta_min}, one thus obtains
\begin{equation}\label{eq:two-min}
  \theta_\mathrm{min}\geq\theta_\mathrm{evp-min}(\Delta T),
\end{equation}
which presents a necessary condition for matching of two regions.
With the equality sign $\theta_\mathrm{min}=\theta_{\mathrm{evp-min}}(\Delta T_\mathrm{max})$, this equation defines a superheating limit $\Delta T_\mathrm{max}$. Since $\theta_{\mathrm{evp-min}}(\Delta T)$ is an increasing function (cf. Fig.~\ref{fig:ThetaApp-dT}), the superheating limit $\Delta T_\mathrm{max}$ is an upper bound. Thus the inequality~\eqref{eq:two-min} can hold when $\Delta T < \Delta T_\mathrm{max}$.  To determine graphically $\Delta T_\mathrm{max}$, Fig.~\ref{fig:theta_app_c-vs-deltaT-and-theta_app-vs-t} presents an example where $\theta(t)$ (bottom and vertical axes, extracted from Fig.~\ref{fig:theta_app-vs-t}) is plotted together with $\theta_{\mathrm{evp-min}}(\Delta T)$ (top and vertical axes, extracted from Fig.~\ref{fig:ThetaApp-dT}). During the oscillation, the minimum value $\theta_\mathrm{min}\simeq 1.8^\circ$ is attained. From the dependence $\theta_\mathrm{evp-min}(\Delta T)$, one can deduce that $\Delta T_\mathrm{max}\simeq 1$~mK. So the solution of the oscillation problem with evaporation is non-existent if $\Delta T>1$mK.

The reason for this paradox is the pinned contact line: for the receding contact line, another degree of freedom appears so the contact angle is not constrained any more. The contact line is necessarily depinned when $\theta$ attains $\theta \left(\Delta T, 0 \right)$ (actually, a larger value $\theta \left(\Delta T, \theta_\textrm{rec} \right)$ but $\theta \left(\Delta T, 0 \right)$ gives a lower bound).

Note that $\theta_\mathrm{min}$ grows slightly with the meniscus velocity amplitude $Ca_0$ (Fig.~\ref{CA_min_vs_Ca}), so does $\Delta T_\mathrm{max}$. However, the $\Delta T_\mathrm{max}$ value remains of the order of mK. As this maximum superheating is considerably smaller than that encountered in practice (where it is rather of several degrees K, see e.g. \cite{LauraATE17}), one can deduce that the contact line receding at evaporation must be accounted for when the meniscus oscillates.

While the calculation has been carried out here only for the pentane case, one can safely state that $\Delta T_\mathrm{max}$ is much smaller than realistic superheating used in thermal engineering applications for many other fluids.

\section{Conclusions}

We have analysed the liquid film deposited by an oscillating liquid meniscus in a capillary tube for the case where the contact line is pinned at the farthest meniscus position. The liquid film thickness is not homogeneous because of the varying meniscus velocity. The periodic solution for such a problem has been identified. The average film thickness is one of the most important quantities. It has been analysed depending on the main system parameters, which are the initial contact angle (that at the farthest meniscus position), the period of oscillation and the amplitude of the meniscus velocity represented with the dimensionless capillary number. The average film thickness depends only weakly on the initial contact angle and grows with the oscillation period until saturation. Globally, the average film thickness is well described by the quasi-steady approach.

Both the meniscus curvature and the contact angle vary in time during such a motion. The contact angle remains nearly constant for a large part of the period. This constant value is independent of both the initial contact angle and the oscillation period. The minimal contact angle encountered during oscillation turns out to be an important quantity. It occurs when the largest film ripple approaches the contact line during the meniscus advance. The minimal contact angle weakly depends on both the oscillation period and the initial contact angle and grows with the capillary number. Its value remains, however, small, of the order of several degrees.

Understanding of evaporation that occurs simultaneously with oscillation is important for applications. The strongest impact of evaporation concerns the contact line vicinity where the liquid film is the thinnest. It is shown that the minimal contact angle that occurs during oscillation with the pinned contact line sets an upper bound for the tube superheating, for which a solution for such a problem exists. This upper bound is quite small (e.g. it is $\sim 1$~mK for the pentane at 1~bar), much smaller than typical experimental superheating. This shows the necessity of considering the contact line receding during the simultaneous oscillation and evaporation. Such a result is important, e.g. for theoretical modelling of the pulsating heat pipes.

\section*{Declaration of Interests}

The authors report no conflict of interest.

\section*{Acknowledgments}

The present work is supported by the project TOPDESS, financed through the Microgravity Application Program by the European Space Agency. This article is also a part of the PhD thesis of X.~Z. co-financed by the CNES and the CEA NUMERICS program, which has received funding from the European Union Horizon 2020 research and innovation program under the Marie Sklodowska-Curie grant agreement No. 800945. An additional financial support of CNES awarded through GdR MFA is acknowledged. We are indebted to S.~Lips for opening the access to his experimental results.

\appendix
\section{Film relaxation}\label{sec:relax}

The meniscus oscillation results in a continuous film shape variation. For this reason, an important quantity is the relaxation time $t_\mathrm{rel}$. It is a characteristic time scale of decrease of a film perturbation caused by the meniscus velocity change.  On a time scale $\gg t_\mathrm{rel}$, one expects the meniscus to behave as if the velocity were constant at each time moment (i.e. in a quasi-steady way).

\begin{figure}
  \centering
  \subfigure[ ]
  {\includegraphics[width=0.48\textwidth]{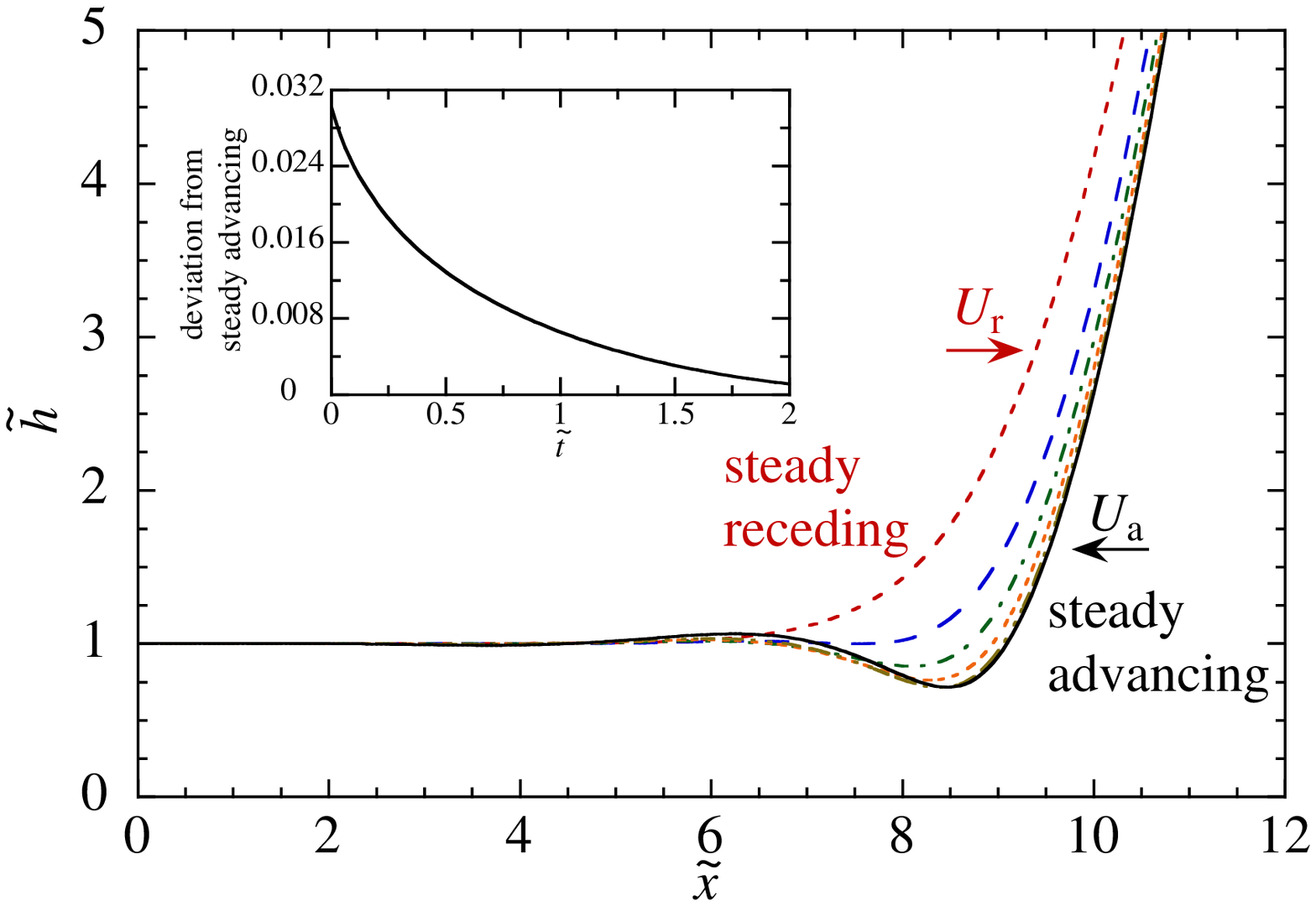}\label{fig:film-rlx}}  \hfill
  \subfigure[]
  {\includegraphics[width=0.48\textwidth]{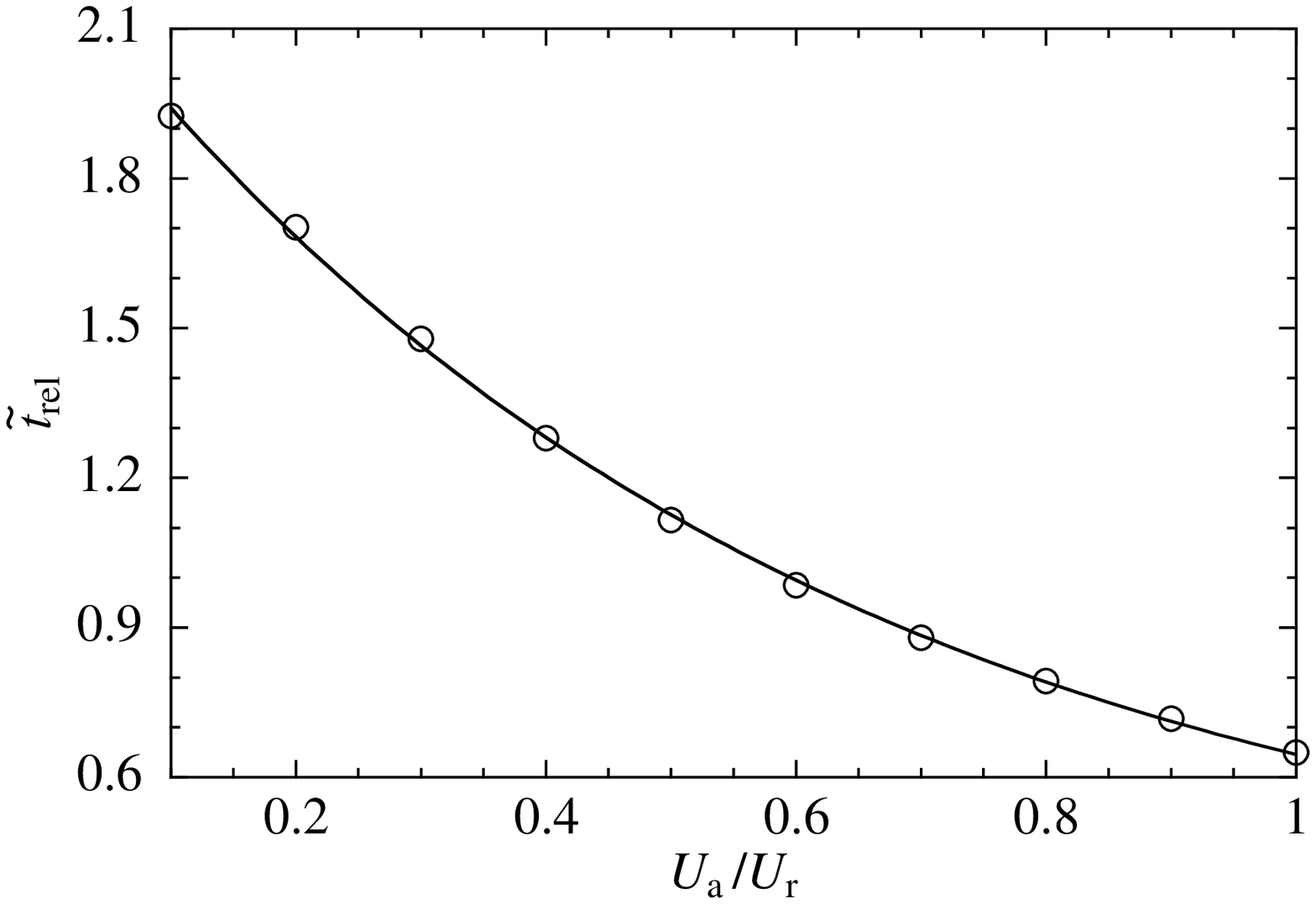}\label{fig:trel_vs_U}}
  \caption{Film relaxation: profile evolution and relaxation time. (a) Film profile evolution (shown in the frame of reference of the meniscus) from steady receding with the velocity $U_\mathrm{r}$ to steady advance with the velocity $U_\mathrm{a}$ such that $|U_\mathrm{a}|=U_\mathrm{r}$. The time lag between the two curves is $\Delta\tilde{t} = 0.5$. The inset shows the root mean square deviation from the steady advancing profile and (b) relaxation time as a function of $U_\mathrm{a}/U_\mathrm{r} $. The dots are the numerical points. The line is the exponential fit.}\label{fig:rlx}
\end{figure}

Relaxation of film profile is studied here on an example of a sudden change in the meniscus motion direction, from receding to advancing. The initial profile (red monotonic curve in Fig.~\ref{fig:film-rlx}) is that of \citet{LL42} defined by Eq.~\eqref{eq:LL} describing the meniscus receding at a constant velocity $U_\mathrm{r}$. In this calculation, $R_\mathrm{m}$ is assumed to remain constant ($=R$) because $Ca_r$ is small. The calculation is performed for $Ca_r=10^{-3}$ but the results are independent of $Ca_r$ provided it is small enough. The deposited film thickness is given by Eq.~\eqref{Breth}.

At $t=0$, the meniscus makes a sudden change of its motion direction and for $t>0$ advances over the liquid film at a constant velocity $U_\mathrm{a}$. The film is infinite so the meniscus frame of reference and thus Eq.~\eqref{eq:GEm} with $U=-U_\mathrm{a}<0$ are employed. With the scaling of Table~\ref{tab:references}, the dimensionless governing equation is
\begin{equation}
\label{eq:GEm-dimensionless-flx}
\frac{\partial \tilde{h}}{\partial \tilde{t}} + \frac{\partial }{{\partial \tilde x}}\left( {\frac{{\tilde h}^3}{3} \frac{{{\partial ^3}\tilde h}}{{\partial {\tilde x ^3}}}} - \tilde U \tilde h \right) =0,
\end{equation}
where $\tilde{U}=-U_\mathrm{a}/U_\mathrm{r}$. The boundary conditions are Eqs.~(\ref{eq:bc_LLl}, \ref{eq:bc_LLr}), where both $\alpha$ and $h_\mathrm{r}$ are now known.

Eq.~\eqref{eq:GEm-dimensionless-flx} is solved numerically, cf. sec.~\ref{NumSec} for the details.

The time evolution of the interface profile is shown in Fig.~\ref{fig:film-rlx}. At $\tilde t=0$, the profile is given by the \citeauthor{LL42} profile. As the time increases, the film relaxes to the steady advancing profile (the solid curve in Fig.~\ref{fig:film-rlx}), which coincides with the \citeauthor{Bretherton} rear meniscus profile. By fitting the root mean square deviation (shown in the inset) from the steady advancing profile, one can introduce the relaxation time $t_\mathrm{rel}$; for $U_\mathrm{a}/U_\mathrm{r}=1$, $\tilde t_\mathrm{rel} \simeq 0.65$.

The calculation shows (Fig.~\ref{fig:trel_vs_U}) that the relaxation time $t_\mathrm{rel}$ decreases with $U_\mathrm{a}$ and remains smaller than 2. As the exponential vanishes after $\tilde t\simeq 5\tilde t_\mathrm{rel}$, one expects that the system behaves independently of a current state after a time lag $\tilde t\simeq 10$. This means that, for $\tilde{P}>100$ considered above, the transient evolution needs to be considered only in the very beginning and the very end of a period where the contact line affects the overall interface shape. For the remaining part of a period, a quasi-static approach is expected to be valid.

\section{Stokes problem of the straight wedge with a varying angle}\label{MoffatSec}

\begin{figure}
  \centering
  \includegraphics[width=0.65\textwidth, clip]{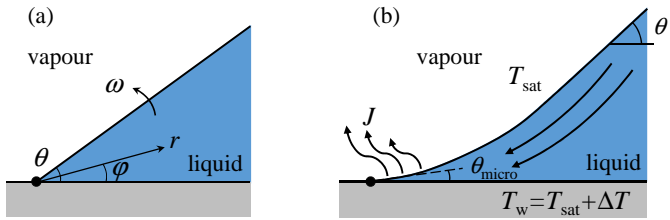}
  \caption{Sketch of the straight liquid wedges with varying angle (a) and curved wedge at evaporation (b).}\label{fig:contact-angles}
\end{figure}

We search a solution of the Stokes problem inside a straight two-dimensional liquid wedge (Fig.~\ref{fig:contact-angles}a) where the opening angle $\theta$ varies with the angular velocity $\omega$. In polar coordinates ($r, \varphi$), the Stokes equations for the liquid velocity $\vec v=(v_r,\,v_\varphi)$ read
\begin{subequations}\label{Stokes}
\begin{align}
  \frac{\partial p_l}{\partial r}=& \mu\left\{\frac{\partial }{\partial r}\left[\frac{1}{r}\frac{\partial (rv_r)}{\partial r}\right]+\frac{1}{r^2}\frac{\partial^2 v_r}{\partial \varphi^2}-\frac{2}{r^2}\frac{\partial v_\varphi}{\partial \varphi}\right\} \label{Stokesr}\\
  \frac{\partial p_l}{\partial \varphi}=& r\mu\left\{\frac{\partial }{\partial r}\left[\frac{1}{r}\frac{\partial (rv_\varphi)}{\partial r}\right]+\frac{1}{r^2}\frac{\partial^2 v_\varphi}{\partial \varphi^2}+\frac{2}{r^2}\frac{\partial v_r}{\partial \varphi}\right\}\label{Stokesphi}\\
  \nabla\cdot\vec v=&0.
\end{align}
\end{subequations}
It is well known that this is equivalent to the streamfunction formulation,
\begin{equation}\label{stream}
\nabla^4\psi=0,
\end{equation}
where
\begin{align}
  v_r & =\frac{1}{r}\frac{\partial \psi}{\partial \varphi} \label{vr}\\
  v_\varphi & =-\frac{\partial \psi}{\partial r} .\label{vphi}
\end{align}
Eq.~\eqref{stream} admits a solution $\psi\propto r^\lambda f(\varphi)$, where $\lambda$ is a constant \citep{Moffatt64}. On the one hand, $\psi$ should scale with $\omega$ because it causes the flow. On the other, from Eq.~\eqref{vphi}, its dimension should be length$^2$/time. The only choice is thus
\begin{equation}\label{psiForm}
\psi=\omega r^2f(\varphi).
\end{equation}
The corresponding function $f$ is $f(\varphi)=A\cos 2\varphi+B\sin 2\varphi+C\varphi+D$ \citep{Moffatt64},
where $A,B,C,D$ are constants. They can be determined with the boundary conditions. A zero velocity ($v_r=v_\varphi=0$) is imposed at the liquid--solid surface $\varphi=0$. A radial velocity $v_\varphi=\omega r$ is imposed at the liquid--vapour boundary $\varphi=\theta$, which is also stress free ($\partial v_r / \partial\varphi=0$). The result is
\begin{equation}\label{ffunc}
\psi=\omega r^2\frac{\tan 2\theta(\cos 2\varphi-1)+2\varphi-\sin 2\varphi}{2(\tan 2\theta-2\theta)}.
\end{equation}
By using it in Eqs.~\eqref{Stokes}, one gets
\begin{align}
\frac{\partial p_l}{\partial r}=&\frac{4\mu\omega }{r(\tan 2\theta-2\theta)},\label{plr} \\
 \frac{\partial p_l}{\partial \varphi}=&0,\nonumber
\end{align}
so $p_l\sim\log r$. Note that for a small $\theta$, Eq.~\eqref{plr} reduces to
\begin{equation}\label{plrl}
\frac{\partial p_l}{\partial r}=\frac{3\mu\omega }{2r\theta^3}.
\end{equation}

\section{Lubrication approach to the evaporation problem}\label{LubAsSec}

\subsection{Straight wedge flow}

Consider first the case with no phase change. It is described by Eq.~\eqref{eq:dimensional-GE}, where $J=0$. As in the above Stokes problem (Appendix~\ref{MoffatSec}), one can consider asymptotically (near the contact line) the straight wedge $h=\theta x$ with the angular velocity $\partial\theta/\partial t=\omega$. The governing equation
\begin{equation*}
\mu\omega x+\frac{\partial}{\partial x}\left( \frac{\theta^3 x^3}{3}\frac{\partial \Delta p}{\partial x} \right) =0
\end{equation*}
results in
\begin{equation}\label{Dpxl}
\frac{\partial \Delta p}{\partial x}=-\frac{3\mu\omega}{2x\theta^3}.
\end{equation}
With no surprise, this expression agrees with the small $\theta$ asymptotics \eqref{plrl} of the Stokes approach and leads to the logarithmic pressure divergence at the contact line.

Consider now the lubrication theory for the volatile liquids that accounts for the Kelvin effect, Eq.~\eqref{eq:GE-kelvin}. For the straight wedge with the varying contact angle it becomes
\begin{equation}\label{GEas1}
\mu\omega x + \frac{\partial }{\partial x}\left(\frac{\theta^3 x^3}{3} \frac{\partial\Delta p}{\partial x}\right)=\frac{\Delta p-\Delta p_\mathrm{cl}}{\theta x}\frac{\mu kT_\mathrm{sat}}{ ({\cal L} \rho)^2}.
\end{equation}
For the fixed contact angle case ($\omega=0$), this equation admits an analytical solution \citep{EuLet12} that satisfies the condition \eqref{eq:p_cl}:
\begin{equation}\label{Jan}
\Delta p=\Delta p_\mathrm{cl}\left[1-\frac{\ell_\mathrm{K}}{x}K_1\left(\frac{\ell_\mathrm{K}}{x}\right)\right],
\end{equation}
where $K_1(\cdot)$ is the modified Bessel function of the first order and
\begin{equation}\label{lK}
 \ell_\mathrm{K}=\frac{\sqrt{3\mu kT_\mathrm{sat}}}{{\cal L} \rho\theta^2}
\end{equation}
is a characteristic length of the Kelvin effect. At $x\ll\ell_\mathrm{K}$,
\begin{equation}\label{asKel}
\Delta p=\Delta p_\mathrm{cl}\left\{1-\exp\left(-\frac{\ell_\mathrm{K}}{x}\right)\left[\sqrt{\frac{\pi}{2}\frac{\ell_\mathrm{K}}{x}}+{\cal O}(x^{1/2})\right]\right\}.
\end{equation}
For the case of varying contact angle $\omega\neq 0$, a solution that satisfies the condition \eqref{eq:p_cl} can be found as an asymptotic expansion
\begin{equation}\label{powAs}
\Delta p=\Delta p_\mathrm{cl}+\frac{3\mu\omega}{\theta^3\ell_\mathrm{K}^2}x^2+{\cal O}(x^4).
\end{equation}

\subsection{Curved wedge flow caused by evaporation for \texorpdfstring{$\omega=0$}{omega=0}}\label{sec:app}

\begin{figure}
  \centering
  \includegraphics[width=0.5\textwidth]{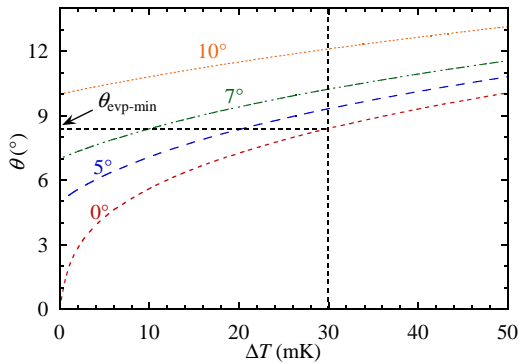}
  \caption{Value of $\theta$ as a function of $\Delta T$ for different $\theta_\mathrm{micro}$ computed for pentane at 1 bar. The curve for $\theta_\mathrm{micro}=0$ corresponds to $\theta_\mathrm{evp-min}(\Delta T)$.}\label{fig:ThetaApp-dT}
\end{figure}
When the substrate is heated, a flow inside the wedge brings the liquid towards the contact line to compensate for the mass loss by evaporation, thus creating the viscous pressure drop described by Eq.~\eqref{Jan}. It can be seen as a curvature that increases when $x\to 0$. The curvature creates a difference between the microscopic contact angle $\theta_\mathrm{micro}$, the actual slope at the contact line, and the interface slope farther away from the contact line, cf. Fig.~\ref{fig:contact-angles}b. The characteristic length for this effect is $\ell_\mathrm{K}\sim 10-100$~nm \citep{PRE13movingCL} from the contact line. At a length scale $x_\mathrm{meso}\gg \ell_\mathrm{K}$ but much smaller than the film-related length scale $\sim 10\,\mu$m, one can define the experimentally measurable interface slope $\theta$, called the apparent contact angle.

The region $x<0<x_\mathrm{meso}$ is often referred to as the microregion. The numerical calculation of  $\theta$ is described by \citet{EuLet12}. It is based on the steady version of Eq.~\eqref{eq:GE-kelvin} (i.e. with $\partial h/\partial t=0$) solved with the boundary conditions (\ref{bc0}, \ref{bc1}). The other two boundary conditions for this fourth-order differential equation are the imposed slope $\theta_\mathrm{micro}$ at the contact line and the condition of zero (on the microregion scale) curvature
\begin{equation}\label{zerocurv}
\left.\frac{\partial^2 h}{\partial x^2}\right|_{x= x_\mathrm{meso}}=0
\end{equation}
Note that the interface slope saturates at $x\gg\ell_\mathrm{K}$ so $\theta$ is independent of $x_\mathrm{meso}$.

Figure~\ref{fig:ThetaApp-dT} demonstrates an example of $\theta$ as a function of $\theta_\mathrm{micro}$ and $\Delta T$ for pentane at 1 bar, cf, Eq.~\eqref{eq:theta_T_theta_mic}. It turns out that $\theta(\Delta T, \theta_\mathrm{micro})$ monotonically grows with both $\theta_\mathrm{micro}$ and intensity of evaporation controlled by $\Delta T$; evidently, $\theta(\Delta T=0)=\theta_\mathrm{micro}$.

One now introduces
\begin{equation}\label{eq:shp-min}
  \theta_\mathrm{evp-min}(\Delta T)\equiv \theta(\Delta T, \theta_\mathrm{micro}=0),
\end{equation}
which is the lower bound of $\theta(\Delta T, \theta_\mathrm{micro})$, i.e. its value for the complete wetting case \citep{PRE13movingCL}. Therefore, the $\theta_\mathrm{micro}=0$ curve in Fig.~\ref{fig:ThetaApp-dT} represents $\theta_\mathrm{evp-min}(\Delta T)$. For example, for $\Delta T=$30~mK, one finds $\theta_\mathrm{evp-min}(\Delta T=30\,\mathrm{mK})\simeq 8.4^\circ$, see Fig.~\ref{fig:ThetaApp-dT}. This means that, for $\Delta T=$30~mK, it is impossible to have an apparent angle smaller than $8.4^\circ$.

\bibliographystyle{jfm}
\bibliography{PHP,Books,Taylor_bubbles,ContactTransf,Hysteresis}

\begin{thebibliography}{32}
\expandafter\ifx\csname natexlab\endcsname\relax\def\natexlab#1{#1}\fi
\def\au#1{#1} \def\ed#1{#1} \def\yr#1{#1}\def\at#1{#1}\def\jt#1{\textit{#1}}
  \def\bt#1{#1}\def\bvol#1{\textbf{#1}} \def\vol#1{#1} \def\pg#1{#1}
  \def\publ#1{#1}\def\arxiv#1{#1}\def\org#1{#1}\def\st#1{\textit{#1}}

\bibitem[Angeli \& Gavriilidis(2008)]{Angeli08}
{\sc \au{Angeli, P.} \& \au{Gavriilidis, A.}} \yr{2008}  \at{Hydrodynamics of
  {T}aylor flow in small channels: A review}.  \jt{Proc. Inst. Mech. Eng. Part
  C J. Mech. Eng. Sci.}  \bvol{222}~(5),  \pg{737 -- 751}.

\bibitem[Aussillous \& Qu\'er\'e(2000)]{Aussillous}
{\sc \au{Aussillous, P.} \& \au{Qu\'er\'e, D.}} \yr{2000}  \at{Quick deposition
  of a fluid on the wall of a tube}.  \jt{Phys. Fluids}  \bvol{12}~(10),
  \pg{2367 -- 2371}.

\bibitem[Baudoin {\em et~al.\/}(2013)Baudoin, Song, Manneville \&
  Baroud]{Baudoin13}
{\sc \au{Baudoin, M.}, \au{Song, Y.}, \au{Manneville, P.} \& \au{Baroud,
  C.~N.}} \yr{2013}  \at{Airway reopening through catastrophic events in a
  hierarchical network}.  \jt{Proc. Natl. Acad. Sci. USA}  \bvol{110}~(3),
  \pg{859 -- 864}.

\bibitem[Bretherton(1961)]{Bretherton}
{\sc \au{Bretherton, F.~P.}} \yr{1961}  \at{The motion of long bubbles in
  tubes}.  \jt{J. Fluid Mech.}  \bvol{10},  \pg{166 -- 188}.

\bibitem[Cherukumudi {\em et~al.\/}(2015)Cherukumudi, Klaseboer, Khan \&
  Manica]{Cherukumudi15}
{\sc \au{Cherukumudi, A.}, \au{Klaseboer, E.}, \au{Khan, S.~A.} \& \au{Manica,
  R.}} \yr{2015}  \at{Prediction of the shape and pressure drop of {T}aylor
  bubbles in circular tubes}.  \jt{Microfluid. Nanofluid.}  \bvol{19}~(5),
  \pg{1221 -- 1233}.

\bibitem[Fourgeaud {\em et~al.\/}(2016)Fourgeaud, Ercolani, Duplat, Gully \&
  Nikolayev]{PRF16}
{\sc \au{Fourgeaud, L.}, \au{Ercolani, E.}, \au{Duplat, J.}, \au{Gully, P.} \&
  \au{Nikolayev, V.~S.}} \yr{2016}  \at{Evaporation-driven dewetting of a
  liquid film}.  \jt{Phys. Rev. Fluids}  \bvol{1}~(4),  \pg{041901}.

\bibitem[Fourgeaud {\em et~al.\/}(2017)Fourgeaud, Nikolayev, Ercolani, Duplat
  \& Gully]{LauraATE17}
{\sc \au{Fourgeaud, L.}, \au{Nikolayev, V.~S.}, \au{Ercolani, E.}, \au{Duplat,
  J.} \& \au{Gully, P.}} \yr{2017}  \at{In situ investigation of liquid films
  in pulsating heat pipe}.  \jt{Appl. Therm. Eng.}  \bvol{126},  \pg{1023 --
  1028}.

\bibitem[de~Gennes(1985)]{deG}
{\sc \au{de~Gennes, P.-G.}} \yr{1985}  \at{Wetting: statics and dynamics}.
  \jt{Rev. Mod. Phys.}  \bvol{57},  \pg{827 -- 863}.

\bibitem[Iliev {\em et~al.\/}(2014)Iliev, Pesheva \& Nikolayev]{PRE14}
{\sc \au{Iliev, S.}, \au{Pesheva, N.} \& \au{Nikolayev, V.~S.}} \yr{2014}
  \at{Contact angle hysteresis and pinning at periodic defects in statics}.
  \jt{Phys. Rev. E}  \bvol{90},  \pg{012406}.

\bibitem[Jane\v{c}ek {\em et~al.\/}(2013)Jane\v{c}ek, Andreotti, Pra\v{z}\'ak,
  B\'arta \& Nikolayev]{PRE13movingCL}
{\sc \au{Jane\v{c}ek, V.}, \au{Andreotti, B.}, \au{Pra\v{z}\'ak, D.},
  \au{B\'arta, T.} \& \au{Nikolayev, V.~S.}} \yr{2013}  \at{Moving contact line
  of a volatile fluid}.  \jt{Phys. Rev. E}  \bvol{88}~(6),  \pg{060404}.

\bibitem[Jane\v{c}ek \& Nikolayev(2012)]{EuLet12}
{\sc \au{Jane\v{c}ek, V.} \& \au{Nikolayev, V.~S.}} \yr{2012}  \at{Contact line
  singularity at partial wetting during evaporation driven by substrate
  heating}.  \jt{Europhys. Lett.}  \bvol{100}~(1),  \pg{14003}.

\bibitem[Klaseboer {\em et~al.\/}(2014)Klaseboer, Gupta \& Manica]{Klaseboer14}
{\sc \au{Klaseboer, E.}, \au{Gupta, R.} \& \au{Manica, R.}} \yr{2014}  \at{An
  extended {B}retherton model for long {T}aylor bubbles at moderate capillary
  numbers}.  \jt{Phys. Fluids}  \bvol{26}~(3),  \pg{032107}.

\bibitem[Landau \& Levich(1942)]{LL42}
{\sc \au{Landau, L.~D.} \& \au{Levich, B.~V.}} \yr{1942}  \at{Dragging of a
  liquid by a moving plate}.  \jt{Acta physico-chimica {USSR}}  \bvol{17},
  \pg{42 -- 54}.

\bibitem[Launay {\em et~al.\/}(2007)Launay, Platel, Dutour \& Joly]{Launay07}
{\sc \au{Launay, S.}, \au{Platel, V.}, \au{Dutour, S.} \& \au{Joly, J.-L.}}
  \yr{2007}  \at{Transient modeling of loop heat pipes for the oscillating
  behavior study}.  \jt{J. Thermophys. Heat Transfer}  \bvol{21}~(3),  \pg{487
  -- 495}.

\bibitem[Lips {\em et~al.\/}(2010)Lips, Bensalem, Bertin, Ayel, Romestant \&
  Bonjour]{Lips10}
{\sc \au{Lips, S.}, \au{Bensalem, A.}, \au{Bertin, Y.}, \au{Ayel, V.},
  \au{Romestant, C.} \& \au{Bonjour, J.}} \yr{2010}  \at{Experimental evidences
  of distinct heat transfer regimes in pulsating heat pipes ({PHP})}.
  \jt{Appl. Therm. Eng.}  \bvol{30}~(8-9),  \pg{900 -- 907}.

\bibitem[Maleki {\em et~al.\/}(2011)Maleki, Reyssat, Restagno, Qu\'er\'e \&
  Clanet]{Maleki11}
{\sc \au{Maleki, M.}, \au{Reyssat, M.}, \au{Restagno, F.}, \au{Qu\'er\'e, D.}
  \& \au{Clanet, C.}} \yr{2011}  \at{{L}andau-{L}evich menisci}.  \jt{J.
  Colloid Interface Sci.}  \bvol{354}~(1),  \pg{359 -- 363}.

\bibitem[Marengo \& Nikolayev(2018)]{EncycExp18}
{\sc \au{Marengo, M.} \& \au{Nikolayev, V.}} \yr{2018}  \at{Pulsating heat
  pipes: Experimental analysis, design and applications}.  \bt{In {\em
  Encyclopedia of Two-Phase Heat Transfer and Flow {IV}\/} (ed. \ed{J.~R.
  Thome})}, ,  \vol{vol. 1: Modeling of Two-Phase Flows and Heat Transfer},
  \pg{pp. 1 -- 62}.  \publ{World Scientific}.

\bibitem[Moffatt(1964)]{Moffatt64}
{\sc \au{Moffatt, H.~K.}} \yr{1964}  \at{Viscous and resistive eddies near a
  sharp corner}.  \jt{J. Fluid Mech.}  \bvol{18}~(1),  \pg{1 -- 18}.

\bibitem[Mohammadi \& Sharp(2015)]{Mohammadi15}
{\sc \au{Mohammadi, M.} \& \au{Sharp, K.~V.}} \yr{2015}  \at{The role of
  contact line (pinning) forces on bubble blockage in microchannels}.  \jt{J.
  Fluids Eng.}  \bvol{137}~(3),  \pg{031208}.

\bibitem[Mortagne {\em et~al.\/}(2017)Mortagne, Lippera, Tordjeman, Benzaquen
  \& Ondar\c{c}uhu]{Mortagne17}
{\sc \au{Mortagne, C.}, \au{Lippera, K.}, \au{Tordjeman, P.}, \au{Benzaquen,
  M.} \& \au{Ondar\c{c}uhu, T.}} \yr{2017}  \at{Dynamics of anchored
  oscillating nanomenisci}.  \jt{Phys. Rev. Fluids}  \bvol{2},  \pg{102201}.

\bibitem[Nikolayev(2010)]{PF10}
{\sc \au{Nikolayev, V.~S.}} \yr{2010}  \at{Dynamics of the triple contact line
  on a nonisothermal heater at partial wetting}.  \jt{Phys. Fluids}
  \bvol{22}~(8),  \pg{082105}.

\bibitem[Nikolayev(2021)]{ATE21}
{\sc \au{Nikolayev, V.~S.}} \yr{2021}  \at{Physical principles and
  state-of-the-art of modeling of the pulsating heat pipe: {A} review}.
  \jt{Appl. Therm. Eng.}  \bvol{195},  \pg{117111}.

\bibitem[Nikolayev \& Sundararaj(2014)]{Nikolayev14}
{\sc \au{Nikolayev, V.~S.} \& \au{Sundararaj, S.}} \yr{2014}  \at{Oscillating
  menisci and liquid films at evaporation/condensation}.  \jt{Heat Pipe Sci.
  Technol.}  \bvol{5}~(1-4),  \pg{59 -- 67}.

\bibitem[Patankar(1980)]{Patankar}
{\sc \au{Patankar, S.~V.}} \yr{1980} {\em Numerical heat transfer and fluid
  flow\/}.  \publ{Washington: Hemisphere}.

\bibitem[Rao {\em et~al.\/}(2017)Rao, Lef\`evre, Czujko, Khandekar \&
  Bonjour]{Rao17}
{\sc \au{Rao, M.}, \au{Lef\`evre, F.}, \au{Czujko, P.-C.}, \au{Khandekar, S.}
  \& \au{Bonjour, J.}} \yr{2017}  \at{Numerical and experimental investigations
  of thermally induced oscillating flow inside a capillary tube}.  \jt{Int. J.
  Therm. Sci.}  \bvol{115},  \pg{29 -- 42}.

\bibitem[Savva {\em et~al.\/}(2017)Savva, Rednikov \& Colinet]{Savva17}
{\sc \au{Savva, N.}, \au{Rednikov, A.} \& \au{Colinet, P.}} \yr{2017}
  \at{Asymptotic analysis of the evaporation dynamics of partially wetting
  droplets}.  \jt{J. Fluid Mech.}  \bvol{824},  \pg{574 -- 623}.

\bibitem[Sign\'e~Mamba {\em et~al.\/}(2018)Sign\'e~Mamba, Magniez, Zoueshtiagh
  \& Baudoin]{Mamba18}
{\sc \au{Sign\'e~Mamba, S.}, \au{Magniez, J.~C.}, \au{Zoueshtiagh, F.} \&
  \au{Baudoin, M.}} \yr{2018}  \at{Dynamics of a liquid plug in a capillary
  tube under cyclic forcing: memory effects and airway reopening}.  \jt{J.
  Fluid Mech.}  \bvol{838},  \pg{165 -- 191}.

\bibitem[Talimi {\em et~al.\/}(2012)Talimi, Muzychka \& Kocabiyik]{Talimi12}
{\sc \au{Talimi, V.}, \au{Muzychka, Y.~S.} \& \au{Kocabiyik, S.}} \yr{2012}
  \at{A review on numerical studies of slug flow hydrodynamics and heat
  transfer in microtubes and microchannels}.  \jt{Int. J. Multiphase Flow}
  \bvol{39},  \pg{88 -- 104}.

\bibitem[Taylor(1961)]{Taylor}
{\sc \au{Taylor, G.~I.}} \yr{1961}  \at{Deposition of a viscous fluid on the
  wall of a tube}.  \jt{J. Fluid Mech.}  \bvol{10}~(2),  \pg{161 -- 165}.

\bibitem[Ting \& Perlin(1987)]{Ting}
{\sc \au{Ting, C.-L.} \& \au{Perlin, M.}} \yr{1987}  \at{Boundary conditions in
  the vicinity of the contact line at a vertically oscillating upright plate:
  an experimental investigation}.  \jt{J. Fluid Mech.}  \bvol{179},  \pg{253 --
  266}.

\bibitem[Youn {\em et~al.\/}(2018)Youn, Han \& Shikazono]{Youn18}
{\sc \au{Youn, Y.~J.}, \au{Han, Y.} \& \au{Shikazono, N.}} \yr{2018}
  \at{Liquid film thicknesses of oscillating slug flows in a capillary tube}.
  \jt{Int. J. Heat Mass Transfer}  \bvol{124},  \pg{543 -- 551}.

\bibitem[Zhang {\em et~al.\/}(1998)Zhang, Hou \& Sun]{Zhang98}
{\sc \au{Zhang, J.}, \au{Hou, Z.} \& \au{Sun, C.}} \yr{1998}  \at{Theoretical
  analysis of the pressure oscillation phenomena in capillary pumped loop}.
  \jt{J. Therm. Sci.}  \bvol{7}~(2),  \pg{89 -- 96}.

\end{thebibliography}

\end{document}